\documentclass[%
 aip,
 amsmath,amssymb,
 reprint,%
]{revtex4-2}

\usepackage{graphicx}
 \graphicspath{{figures/}}
\draft % marks overfull lines with a black rule on the right
\usepackage{amsmath}
\usepackage{siunitx}
\usepackage{float}
\makeatletter
\let\newfloat\newfloat@ltx
\makeatother
\usepackage{algorithm}
\usepackage{algpseudocode}

\algrenewcommand\algorithmicrequire{\textbf{Input:}}
\algrenewcommand\algorithmicensure{\textbf{Output:}}

\begin{document}

%\preprint{}

\title{Continuum kinetic investigation of the impact of bias potentials in the current saturation regime on sheath formation}



\author{C. R. Skolar}   %Am I allowed to do this to my name?
\email[]{chiragr@vt.edu}
%\homepage[]{https://www.aoe.vt.edu/people/faculty/srinivasan/personal-page.html}
%\thanks{}
\affiliation{Kevin T. Crofton Department of Aerospace and Ocean Engineering, Virginia Tech, Blacksburg, VA 24060, USA}

\author{K. Bradshaw}
\email[]{kolterb@vt.edu}
%\homepage[]{https://www.aoe.vt.edu/people/faculty/srinivasan/personal-page.html}
%\thanks{}
\affiliation{Kevin T. Crofton Department of Aerospace and Ocean Engineering, Virginia Tech, Blacksburg, VA 24060, USA}

\author{J. Juno}
\email[]{jjuno@pppl.gov}
%\homepage[]{https://www.aoe.vt.edu/people/faculty/srinivasan/personal-page.html}
%\thanks{}
\affiliation{Princeton Plasma Physics Laboratory, Princeton, NJ 08540, USA}

\author{B. Srinivasan}
\email[]{srinbhu@vt.edu}
%\homepage[]{https://www.aoe.vt.edu/people/faculty/srinivasan/personal-page.html}
%\thanks{}
\affiliation{Kevin T. Crofton Department of Aerospace and Ocean Engineering, Virginia Tech, Blacksburg, VA 24060, USA}

% Anyone else? Dunno about Jimmy anymore.
% Maybe Kolter goes in here? I dunno. He helped me a lot with input files and such. If not here, at least an acknowledgement
% Petr? I will be using a version of his Robertson solver that I modified


\date{\today}

% revisit and make sound nicer
\begin{abstract}
	In this work, we examine sheath formation in the presence of bias potentials in the current saturation regime for pulsed power fusion experiments. It is important to understand how the particle and heat fluxes at the wall may impact the wall material and affect electrode degradation. Simulations are performed using the 1X-1V Boltzmann-Poisson system for a proton-electron plasma in the presence of bias potentials ranging from 0 to 10 kV. The results indicate that the sheath near the anode remains generally the same as that of a classical sheath without the presence of a bias potential. However, the sheath near the cathode becomes more prominent with a larger potential drop, a significant decrease of electron density, and larger sheath lengths. The spatially constant current density increases to a saturation value with increasing bias potential. For high bias potentials, the material choice needs to consider that the anode faces significantly larger particle and heat fluxes compared to the cathode. In general, the results trend with theory with differences attributed to the simplified assumptions in the theory and the kinetic effects considered in the simulations. Due to the significant computational cost of a well resolved 1X-2V simulation, only one such simulation is performed for the 5 kV case showing higher current.
\end{abstract}
%
%	In this work, we examine sheath formation in the presence of bias potentials in the current saturation regime for pulsed power fusion experiments. It is important to understand how the particle and heat fluxes at the wall may impact the wall material and affect electrode degradation. Simulations are performed using the 1X-1V Boltzmann-Poisson system for a proton-electron plasma in the presence of bias potentials ranging from 0 to 10 kV. The results indicate that the sheath near the high potential wall remains generally the same as that of a classical sheath without the presence of a bias potential. However, the sheath near the low potential wall becomes more prominent with a larger potential drop, a significant decrease of electron density, and larger sheath lengths. The spatially constant current density increases to a saturation value with increasing bias potential. The current is dominated by the ions at the low potential wall and by the electrons at the high potential wall. The heat flux increases to a saturation value at the high potential wall and tends to zero at the low potential wall with increasing bias potential. The results trend with theory with differences attributed to the simplified assumptions in the theory and the kinetic effects considered in the simulations. Due to the significant computational cost of a well resolved 1X-2V simulation, only one such simulation is performed for the 5 kV case showing higher current.

%\pacs{}% insert suggested PACS numbers in braces on next line   % What is this?

\maketitle %\maketitle must follow title, authors, abstract and \pacs

\section{Introduction}  
\label{s:intro}

% Should I have individual citations for each of these? Should I add more? 
% What other situations might this be useful? Plasma switches? 
% Plasma guns like what Ameer and Max did? Looks like they do use electrodes and showed a plot of high current with some potential. But is this relevant? Read more
Plasma-material interactions (PMI) are an important area 
of research for any plasma facing surface.
Such examples are ubiquitous in many plasma applications
such as semiconductor etching,
efficiency and wall erosion in fusion experiments
and electric thrusters,
understanding of sensors such as Langmuir probes,
and the impact of space charging on a spacecraft.\cite{stangeby2000plasma,lieberman2005principles,baalrud2020interaction}

When a plasma interacts with a solid wall,
a potential barrier forms that slows electrons and accelerates ions.
This localized region of net positive charge is called the plasma, ion, or Debye sheath;
its length is on the order of a few Debye lengths,
$\lambda_D=\sqrt{\epsilon_0 T_e/ne^2}$.\cite{robertson2013sheaths}
The ion and electron particle fluxes into the wall are equilibrated such
that there is no net current.

However, for a plasma bounded by two walls with an electric potential bias between them,
a net current forms at the walls and within the bulk plasma.\cite{stangeby2000plasma,baalrud2020interaction}
This is a common configuration for many devices such as 
arcjets,\cite{wollenhaupt2018overview}
gas puff Z-pinches,\cite{giuliani2015review},
edge biasing in tokamaks,\cite{weynants1993edge} % MADE FIX HERE
and shear flow stabilized Z-pinches.\cite{shumlak2020z}
For this work, we study sheath formation for bias potentials ranging from 0 to 10 kV, 
well into the current saturation regime
for shear-flow stabilized Z-pinches.
While this work is motivated by the shear-flow stabilized Z-pinch experiments, 
it is generally applicable to plasma sheaths in the presence of biased wall potentials. 

%operate in the regime relevant to shear flow stabilized Z-pinch experiments.

Z-pinches are considered a commercially viable
path to an intermediate-scale thermonuclear fusion
reactor. A Z-pinch forms when a plasma with an 
axial current becomes cylindrically confined by its 
self-generated magnetic field.\cite{hartman1977conceptual,shumlak2020z} The radius of the resulting
column of plasma is called the pinch radius.
In this configuration, increasing the axial, or pinch, current
decreases the pinch radius while increasing the plasma density and 
temperature;\cite{bennett1934magnetically} thus, the probability of nuclear reactions within the plasma 
significantly increases.\cite{shumlak2020z}
These Z-pinch configurations are 
unstable to plasma instabilities, such as the sausage and kink instabilities,\cite{haines2000past}
which cause losses in nuclear fusion energy output.
Experiments such as ZAP\cite{shumlak2001evidence,shumlak2003sheared,shumlak2009equilibrium,shumlak2017increasing} 
and FuZE\cite{zhang2019sustained,mitrani2019measurements,claveau2020plasma,stepanov2020flow} show
that these instabilities are stabilized
by shear flow.\cite{shumlak2012sheared}

The pinch current is generated by two electrodes with a bias potential between them. % Is this the best way to say this?
Understanding how the Z-pinch affects the electrodes, and vice versa, is an open area of research;
this can provide useful insight on electrode degradation. % Is there anything I should cite here?
Depending on ratios of the effective areas of the electrode and the ground wall, different types of sheath behaviors can form, such as 
the plasma or ion sheath, 
the reverse or electron sheath,
the double sheath,
anode glows, 
or fireballs.\cite{baalrud2020interaction}  
Eq.~4 from Ref.~\onlinecite{baalrud2020interaction} is a theoretical
criterion based on effective area ratios between an electrode and a grounding surface
that can determine the type of sheath that develops.
In our case, we assume that the Z-pinch profile at the cathode and anode is the same.
Therefore, based on the criterion, an ion or plasma sheath is expected to develop.
Thus, we can simplify the geometry into one spatial dimension since the plasma sheath is what naturally develops in this approximation.  % This still needs to be said better. Revisit again. Also, do I have anything to cite here? Stangeby book? 

Plasma sheaths have been simulated using 
particle-in-cell (PIC) methods,\cite{scheiner2016particle,li2022bohm,li2022transport} % ALSO ADD YUZHI'S POP HERE. 
fluid methods, \cite{shumlak2011advanced,cagas2017continuum}
and continuum kinetic methods.\cite{cagas2017continuum,cagas2020plasma,bradshaw2022}  
In this paper, we use the continuum kinetic approach for the following reasons.  % Do I put a colon here instead of the period? Maybe rephrase?
Firstly, much of the physics of sheath formation lie at kinetic scales which fluid models cannot resolve.
Secondly, PIC methods produce noisy results compared to continuum kinetic methods.
This can cause difficulties in analyzing the results and makes finding accurate representations of higher order moment
calculations, such as heat flux, more difficult. 
Lastly, continuum kinetic methods can more easily be used for multi-scale hybrid fluid-kinetic solvers. % Is there a good citation for this? I know this is one of the goals that Gkeyll is working toward. Look into this more.

This work aims to model plasma-material interactions in the presence of bias potential walls
using continuum kinetic methods in 1X-1V (one spatial dimension and one velocity dimension) and 1X-2V (one spatial dimension and two velocity dimensions). 
We work in the parameter regime for Z-pinch fusion reactor electrodes, such as FuZE.\cite{zhang2019sustained}
The results presented here provide insight into important wall parameters, such as particle and heat flux, that will inform electrode design. 
Sec.~\ref{s:theory} provides theory for sheath formation with biased walls. 
Sec.~\ref{s:model} discusses the mathematical and numerical framework used to study the sheaths.
Sec.~\ref{s:setup} describes the setup of the sheath simulations.
Sec.~\ref{s:results} presents the results of the simulations with comparison to theory and discussions on how they relate to the broader literature.
Sec.~\ref{s:conclusions} provides the summary and conclusions of this paper.

\section{Theory of Sheaths in the Presence of a Bias Potential}  
\label{s:theory}

Consider an isothermal plasma that is doubly bounded
between two perfectly absorbing walls with no electric potential bias between them. 
If the electrons are assumed to be Boltzmann distributed % MADE FIX HERE
and the ion particle flux is assumed  
to remain approximately constant within the sheath, 
then the plasma potential, 
or the potential in the center of the domain, $\phi_{p_0}$, is
\begin{equation}
	\tilde{\phi}_{p_0} = -\frac{1}{2} \ln 
	\bigg[ \Big( 2 \pi \frac{m_e}{m_i} \Big) 
	\Big(1 + \frac{T_i}{T_e} \Big)\bigg],
	\label{eq:phi_p0}
\end{equation}
where $T$ is the temperature in energetic units.\cite{stangeby2000plasma}  % Should I specific which section of the textbook here?
The tilde above $\phi_{p_0}$ denotes that it is a normalized potential, defined as $\tilde{\phi}=e\phi/T_e$.
Eq.~\ref{eq:phi_p0} assumes that the electric potential at the walls are the ground potential;
therefore, $\tilde{\phi}_{p_0}$ will be positive for realistic values of ion mass. 
In this case, the ion and electron particle fluxes cancel such that there is no current at
the walls and throughout the plasma. This case without a bias potential will hereafter
be referred to as the classical sheath.

Suppose instead there exists an applied bias potential between the two walls, 
or electrodes. 
In this case, we define the higher potential wall (positively-biased wall) 
as the anode and the lower potential wall (negatively-biased wall) as the cathode.
Thus, the potential difference between the anode and the cathode
is defined as $\phi_b = \phi_A - \phi_C$, where
the subscripts $A$ and $C$ denote the anode and cathode, respectively.
If we consider the electric potential at the cathode as the ground potential, 
the anode and cathode potentials become $\phi_A = \phi_b$ and $\phi_C = 0$, respectively.
The resulting plasma potential is\cite{stangeby2000plasma}
\begin{equation}
	\tilde{\phi}_p = \tilde{\phi}_b 
		- \ln \Bigg[ \frac{2 \exp \big( - \tilde{\phi}_{p_0} \big)}{1 + \exp\big( - \tilde{\phi}_b  \big)}   \Bigg].
		\label{eq:phi_p}
\end{equation}

Unlike a classical sheath, a current develops throughout the plasma when a nonzero bias potential exists.
Under the additional assumption that the electrons are Maxwellian with a constant temperature, 
the ion and electron particle fluxes at the electrodes are\cite{stangeby2000plasma}
\begin{align}
	n_iu_i\big|_A &= \frac{1}{2} n_0 c_s 
	\label{eq:niui_lw} \\
	n_eu_e \big|_A &= \frac{1}{8} n_0 \bar{c}_e \exp \big(-\tilde{\phi}_p + \tilde{\phi}_b \big) 
	\label{eq:neue_lw} \\
	n_iu_i\big|_C &= \frac{1}{2} n_0 c_s 
	\label{eq:niui_rw} \\
	n_eu_e \big|_C &= \frac{1}{8} n_0 \bar{c}_e \exp \big(-\tilde{\phi}_p \big),
	\label{eq:neue_rw}
\end{align}
where $c_s=\sqrt{(\gamma_e T_e+ \gamma_i T_i)/m_i}$ is the Bohm speed
and $\bar{c}_e=\sqrt{8T_e/\pi m_e}$ is the mean speed of
the electrons. The ion particle fluxes, Eqs.~\ref{eq:niui_lw} and ~\ref{eq:niui_rw},
are the same for both electrodes because the ions are much more massive than the electrons;
therefore, they are negligibly affected by changing bias potential.
The current density at the electrodes is found 
through clever use of Eq.~\ref{eq:phi_p0} and
either Eqs.~\ref{eq:niui_lw} and~\ref{eq:neue_lw} or
Eqs.~\ref{eq:niui_rw} and~\ref{eq:neue_rw}.
The resulting current density, which is the same at both electrodes, is\cite{stangeby2000plasma} 
\begin{equation}
	j_{A,C} = \frac{1}{2} e n_0 c_s \Big[ 1 - 
	\exp \big( \tilde{\phi}_{p_0} - \tilde{\phi}_{p} \big) \Big].
	\label{eq:j}
\end{equation}
As the bias potential increases, the current saturates to\cite{stangeby2000plasma}
\begin{equation}
	j_{sat} = \lim_{\phi_b \rightarrow\infty} = \frac{1}{2} e n_0 c_s. % MADE FIX HERE
	\label{eq:j_sat}
\end{equation}
Therefore, the saturation current increases with higher background
density and temperature.

\section{Model} \label{s:model}

The plasma is modeled kinetically 
using the Boltzmann-Poisson system
for a two species plasma. The distribution function, $f$, is evolved
through the Boltzmann equation, which is
\begin{equation}
	\frac{\partial f_\alpha}{\partial t}
	+ \nabla_\mathbf{x} \cdot (f_\alpha \mathbf{v})
	+ \frac{q_\alpha}{m_\alpha} \nabla_\mathbf{v} \cdot (f_\alpha \mathbf{E})  
	= \sum \Big( \frac{\partial f_\alpha}{\partial t} \Big)_c + S_\alpha,
	\label{eq:boltzmann}
\end{equation}
where $q$ is the electric charge and $m$ is the mass
for species $\alpha$, ions or electrons. 
The term $\sum(\partial f_\alpha/\partial t)_c$ is
the Dougherty collision operator,\cite{dougherty1964model,francisquez2020conservative,hakim2020conservative} which includes the effects
of all inter- and intra-species collisions. 
The term $S_\alpha$ is a source term to ensure particle 
conservation within the domain.
The source term is of the form from Ref.~\onlinecite{bradshaw2022}
with an included reflection to account for two walls
instead of just one.

For this work, magnetic field effects are not considered;
therefore, the acceleration term in Eq.~\ref{eq:boltzmann}
depends solely on the electric field. The electric field
is calculated using the electrostatic potential, $\phi$,
as $\mathbf{E}=-\nabla \phi$. The electric potential
is found using the Poisson equation,
\begin{equation}
	\nabla^2 \phi = - \frac{e(n_i - n_e)}{\epsilon_0},
	\label{eq:poisson}
\end{equation}
where $n$ is the number density.

Simulating the the entire axial length
of a Z-pinch fusion reactor experiment, 
such as FuZE, is computationally expensive when using kinetic models. 
For perspective, the FuZE assembly region is \SI{50}{\centi \meter}
or almost $5\times10^{5}\lambda_D$,\cite{zhang2019sustained} which is much larger than
can feasibly be simulated with present technology using the continuum kinetic method.
The larger length can be approximated by artificially increasing the collision frequencies,
resulting in smaller mean free paths.
This allows for greater thermalization of the plasma. 
An added bonus is that
the larger collision frequencies also help mitigate any transient
waves launched by the initial conditions. 
The collision frequencies included in the numerical model are
\begin{align}
	\nu_{ee} &= \frac{v_{th_e}}{\lambda_{MFP}} \label{eq:nuee} \\
	\nu_{ei} &= \nu_{ee} \label{eq:nuei}\\
	\nu_{ii} &= \frac{v_{th_i}}{\lambda_{MFP} } \label{eq:nuii}\\
	\nu_{ie} &= \frac{m_e}{m_i} \nu_{ee} \label{eq:nuie} ,
\end{align}
where $\lambda_{MFP}$ is the mean free path
and $v_{th_\alpha}$ is the thermal velocity (defined as
$v_{th_\alpha} = \sqrt{T_\alpha/m_\alpha}$). 
For all simulations performed in this paper, the mean free
path is set to $50\lambda_D$. 
Despite the mean free path being larger than sheath scales,
it is found that using a constant collision frequency in the lower
temperature sheath region results in an effectively smaller mean free path.\cite{bradshaw2022} 
Therefore, we must multiply the collision frequencies
from Eqs.~\ref{eq:nuee} to \ref{eq:nuie} by some function $h(x)$,
defined as 
\begin{equation}
	h(x) = h_0(-x + 128\lambda_D) + h_0(x - L_D + 128 \lambda_D) - 1,
	\label{eq:collProfile}
\end{equation}
where $L_D$ is the domain length and 
\begin{equation}
	h_0(x) = \bigg[  1 + \exp \Big( \frac{x}{12 \lambda_D} - \frac{16}{3} \Big)   \bigg].
	\label{eq:collProfileBase}
\end{equation}
The choice of this profile is somewhat arbitrary, but critically has more collisions in the center of the domain 
sufficient to maintain a Maxwellian presheath, while dropping rapidly as it approaches the wall in order to preserve an 
approximately collisionless sheath condition.

We use Dirichlet boundary conditions for the Poisson equation, Eq.~\ref{eq:poisson}.
For the Boltzmann equation, Eq.~\ref{eq:boltzmann}, we use perfectly absorbing walls
in configuration space and zero gradient in velocity space.

The model outputs are the particle distribution function (from the Boltzmann equation, Eq.~\ref{eq:boltzmann}) for each species (ions and electrons)
and the electric potential (from the Poisson equation, Eq.~\ref{eq:poisson}). 
The bulk properties of the plasma can be found by taking the moments of the distribution function.
The density is the zeroth moment of the distribution function, $n = \int f d\textbf{v}$.
The particle flux is the first moment of the distribution function, $nu_i = \int v_i f d\textbf{v}$.
The temperature in the $x$ direction, can be found from the second moment of the distribution function, 
$M_{2_{ij}}= \int v_i v_j f d \textbf{v}$, as
\begin{equation}
	T_x = m \bigg( \frac{ M_{2_{xx}}}{n} - u_x^2  \bigg).
	\label{eq:T}
\end{equation}
For the $y$ direction temperature, simply replace $x$ with $y$.
For a 2V output, 
the heat flux, $q$, in the $x$ direction
can be calculated from a subset of the third moment
of the distribution function, $M_{3_{iix}} = \int v^2 v_x f d \textbf{v}$, as\cite{wangL2015} 
\begin{multline}   
	q_x = \underbrace{m \Bigg[ \frac{1}{2} M_{3_{iix}} - u_x \bigg( \frac{3}{2} M_{2_{xx}}  - n u_x^2 \bigg)}_{\text{1V}}  \\
	- \frac{1}{2} u_x M_{2_{yy}} - u_y \bigg( M_{2_{xy}} - n u_xu_y \bigg) \Bigg].
\end{multline}
For the 1V simulations, only the part in the underbrace is needed to calculate the heat flux.

\section{Simulation Setup and Numerical Methods}

\label{s:setup}

This section describes the simulation setup for
studying sheaths in the presence of biased potential walls in 1X-1V and 1X-2V.
The Boltzmann equation, Eq.~\ref{eq:boltzmann},
solves the distribution function which uses space and velocity as coordinates. % Revisit how  Isay this 
% MADE FIX HERE

The plasma is modeled as a proton-electron plasma. 
The background density and temperature values are 
$n_0=\SI{1.1e23}{\meter^{-3}}$ and 
$T_0=\SI{2}{\kilo \electronvolt}$, respectively.
These values are based on the upper range
of measured values in FuZE,\cite{zhang2019sustained}
and therefore are expected to yield a higher current and saturation current,
according to Eqs.~\ref{eq:j} and \ref{eq:j_sat}, respectively.

The domain length in configuration space, $L$, is $256\lambda_D$ with 
512 cells. 
The velocity space domain spans $\pm 6 v_{th_\alpha}$ with 64 cells, where
$v_{th_\alpha}$ uses $T_0$ as the temperature, for 
the respective species $\alpha$.

The plasma is initialized isothermally 
such that $T_i=T_e=T_0=\SI{2}{\kilo \electronvolt}$.
The ions and electrons are initialized as Maxwellian distributions,
\begin{equation}
	f_\alpha = 
	\frac{n_{\alpha,0}(x)}{ \big( 2 \pi \big)^{N/2} \prod_{i=1}^{N} v_{th_{i_{\alpha,0}}}}
	\exp \bigg( - \sum_{i=1}^{N} \frac{\big[ v_i - u_{i_{\alpha,0}}(x)  \big]^2}{v_{th_{i_{\alpha}}}^2}  \bigg) ,
	\label{eq:maxwellIC}
\end{equation}
where $n_\alpha(x)$ and
$u_\alpha(x)$ are, respectively, the position
dependent density and drift velocities
for species $\alpha$ and $N$ is the 
number of velocity dimensions considered in the simulation.
For this work, $N$ is either 1 or 2.
Additionally, note that we do not consider a magnetic
field in these simulations. % MADE FIX HERE

\begin{figure}[!htb]
	\centering
	\includegraphics[width=\linewidth]{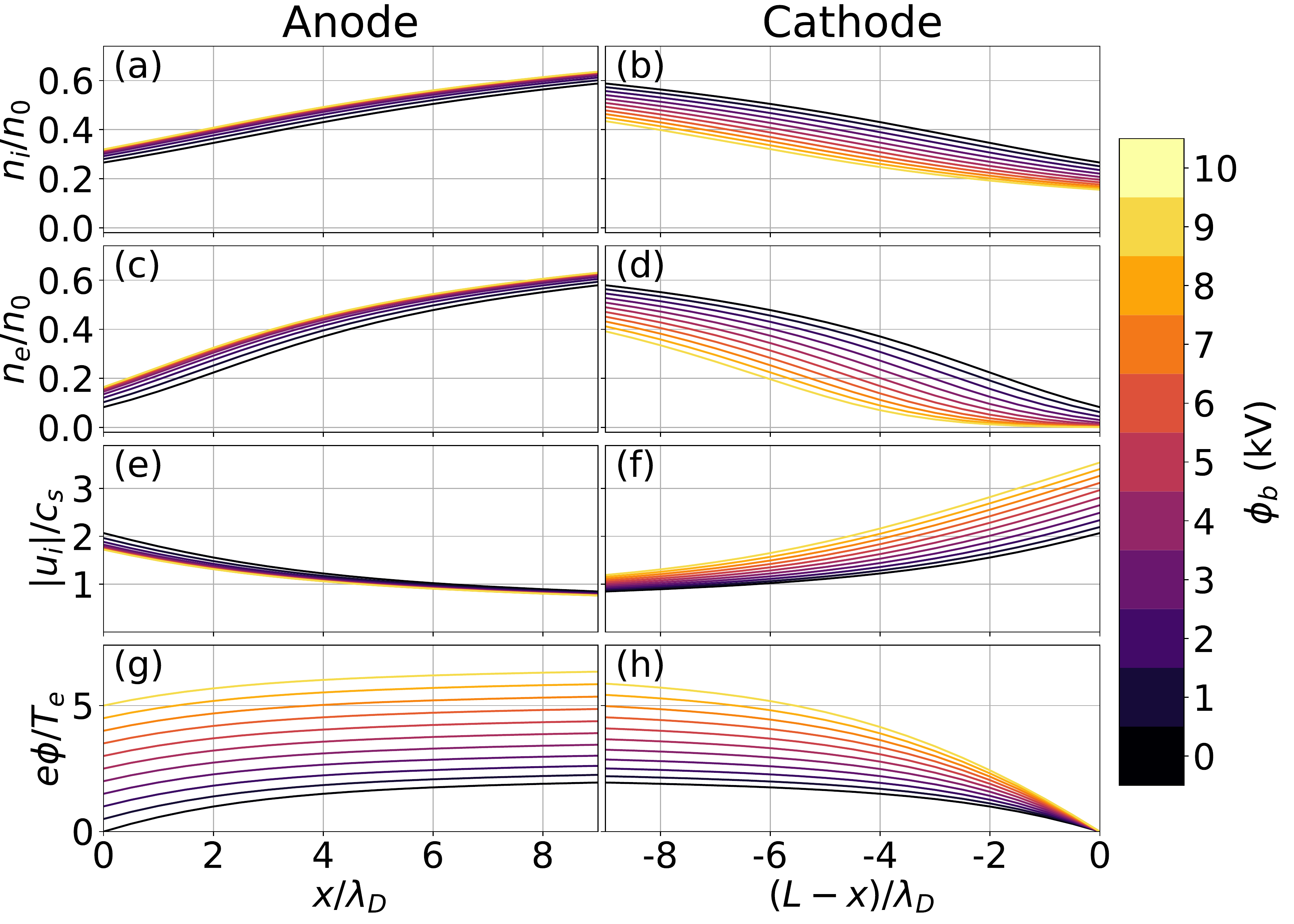}
	\caption{Normalized sheath profiles of the ion density (a,b),
		electron density (c,d),
		ion drift velocity (e,f),
		and electric potential (g,h)
		near the anode and the cathode are
		found by solving Eqs.~\ref{eq:Rob_ODE_phi}-\ref{eq:Rob_ODE_ui}
		for bias potentials of 0 to 10 \si{\kilo \volt}.
		These profiles are used as the initial conditions for the simulations.
		See Appendix~\ref{a:Robertson_solver} for more information
		on how these profiles are calculated.
	}
	\label{f:robertson_profiles}
\end{figure}

Initializing spatially constant densities and drift velocities would
yield the same general quasi-steady state solution. However, the large difference
between the initial conditions and the final solution would launch Langmuir 
waves in the system;\cite{cagas2018continuum}
this would require running to a longer end time to reach quasi-steady state as these waves decay. 
Ref.~\onlinecite{robertson2013sheaths} provides a method  that
calculates approximate solutions that can be used as initial conditions. 
A detailed explanation is found in Appendix~\ref{a:Robertson_solver}.
The electron drift velocity is initialized to be 0, $u_{e,0}(x) = 0$.
The ion density, electron density, and ion drift velocity
are initialized based on the profiles calculated using Eqs.~\ref{eq:Rob_ODE_phi}-\ref{eq:Rob_ODE_ui}.
The initial profiles used in the simulations for bias potentials from 0 to \SI{10}{\kilo \volt} are shown in Fig.~\ref{f:robertson_profiles}.
The regions near the electrodes use the respective profile based
on the theory and calculations from Appendix~\ref{a:Robertson_solver}.   % These last 2 sentences definitely need work
The center plasma region uses constant values such that
the ion and electron density are $n_0=\SI{1.1e23}{\meter^{-3}}$
and the ion drift velocity is \SI{0}{\meter / \second}.
Note how Fig.~\ref{f:robertson_profiles}(f) shows that the ion velocity
at the cathode increases with bias potential. Another utility of using
these initial conditions is that the ion velocity estimate provides
an a priori guess to what the velocity space domain should be for the ions.

% Update the text in this captio nto be more accurate. This is just copy pasted from old location for fig
% Should I show the entire domain or just the sheath regions?

Eqs.~\ref{eq:boltzmann} and \ref{eq:poisson} are solved using 
the code \verb|Gkeyll|.\cite{gkylDocs,juno2018discontinuous,hakim2020alias} % Should I list more?
They are discretized using the discontinuous 
Galerkin method with an orthonormal modal serendipity basis  % Include some references for DG and such. Maybe ask Bhuvana for this one.
with a polynomial order of 2. Eq.~\ref{eq:boltzmann} is 
integrated in time using a 3-stage $3^{\text{rd}}$-order
strong stability preserving Runge-Kutta method.\cite{gottlieb2005high}
Eq.~\ref{eq:poisson} is solved using a direct matrix inversion. % Is this true? Reference to a gkeyll paper that uses this. Look up.

The Dirichlet boundary conditions for the Poisson equation are defined
such that the electric potential at the anode and cathode
are $\phi_b$ and 0, respectively.
1X-1V simulations are performed with bias potentials ranging from 0 to \SI{10}{\kilo \volt}
in increments of \SI{1}{\kilo \volt} based
on observations from FuZE.\cite{stepanov2020flow} % MADE FIX HERE

 In addition, one 1X-2V simulation is performed for 
\SI{5}{\kilo \volt} to obtain an understanding of how the additional velocity dimension 
impacts the solution.

\section{Results and Discussion} \label{s:results}

% Double check this for the 2V cases
All of the results in this section are taken from simulations that
have reached a quasi-steady state at $t=20000\omega_{pe}$
where $\omega_{pe}=\sqrt{n_0e^2/m_e\epsilon_0}$ is the 
plasma frequency.

\subsection{1X-1V Results}

\begin{figure}[!htb]  
	\centering 
	\includegraphics[width=\linewidth]{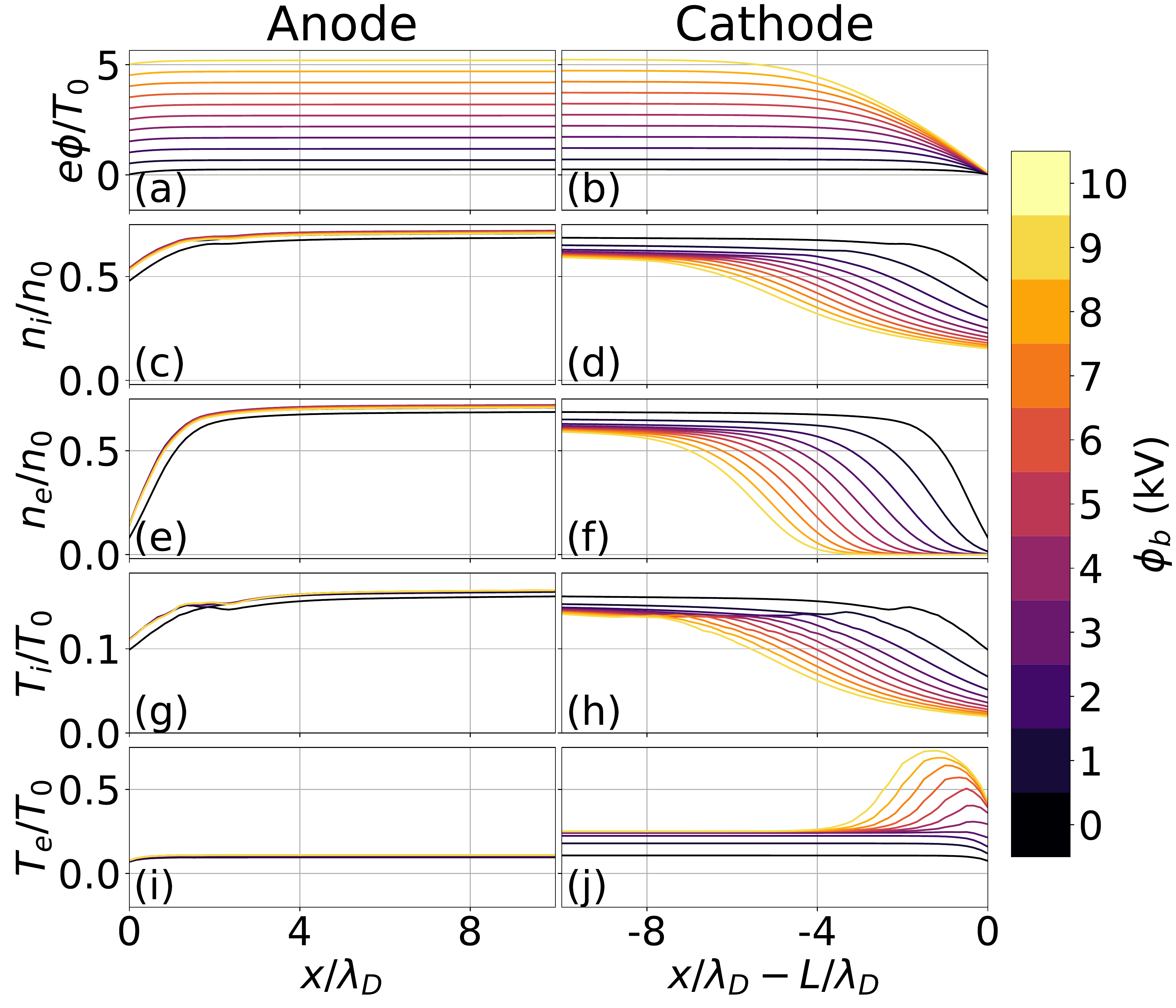}
	\caption{Plots of
		the normalized
		electric potential (a,b),
		ion density (c,d),
		electron density (e,f), 
		ion temperature (g,h), and
		electron temperature (i,j)
		profiles near the anode and cathode
		for bias potentials from 0 to \SI{10}{\kilo \volt}.}
	\label{f:sheath_profiles}
\end{figure}

Fig.~\ref{f:sheath_profiles} shows the profiles of the normalized
electric potential, 
ion density, electron density, 
ion temperature, and electron temperature
in the region of $10\lambda_D$
near the anode and cathode
for varying bias potentials.

The electric potential near the anode, Fig.~\ref{f:sheath_profiles}(a),
retains the same general shape and visually appears to translate upward with increasing bias potential.
In other words, the potential drop at the anode  % MADE FIX HERE
does not change significantly
with bias potential.
Therefore, the ion density,
electron density, 
ion temperature,
and electron temperature at the anode
minimally change with bias potential,
as shown in Figs.~\ref{f:sheath_profiles}(c,e,g,i), respectively.

\begin{figure}[!htb] 
	\centering
	\includegraphics[width=\linewidth]{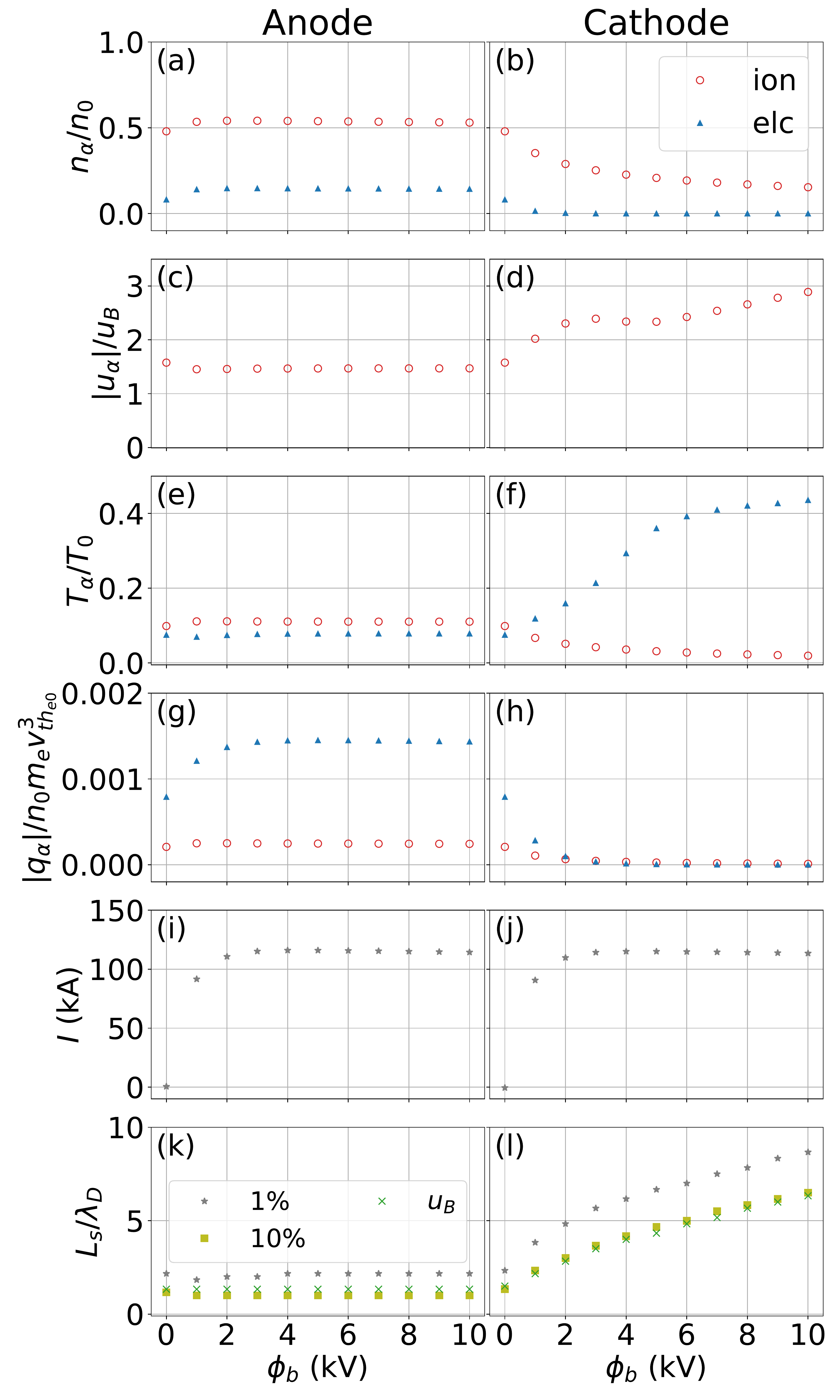}
	\caption{Plots of normalized densities (a,b),
		particle fluxes (c,d),
		temperatures (e,f),
		heat fluxes (g,h),
		current (i,j),
		and sheath lengths (k,l)
		at the electrodes
		as functions of the bias potential. 
		Note that the absolute value of the particle and heat 
		fluxes are plotted; at the anode,
		these are negatively valued. The red circles
		and blue triangles correspond to the ions and 
		electrons, respectively.
		All values except for the current, Panels (i-j), are normalized.
		Note that the drift velocity, Panels (c-d) are normalized by the local
		Bohm speed.
	}
	\label{f:wall_fluxes_SI}
\end{figure}

In contrast,
the electric potential drop near the cathode
gets larger with increasing bias potential, as seen in Fig.~\ref{f:sheath_profiles}(b),
resulting in substantially different profiles for the other variables.
The ion and electron density profiles, Figs.~\ref{f:sheath_profiles}(d,f), respectively,   
show a reduction in density with increasing bias potential. 
In addition, the point at which the densities begin to 
decrease substantially
moves further away from cathode (to the left) with increasing bias potential. 
In other words, as will be shown later in Fig.~\ref{f:wall_fluxes_SI}(l), the 
sheath length increases with bias potential.
The potential barrier is large enough to drive the electron density to 0 at large bias potentials. 
Furthermore, the region in which the electrons have a near zero density becomes larger with increasing bias potential.
The ion temperature, Fig.~\ref{f:sheath_profiles}(h) shows a similar trend to the ion density.
The electron temperature, Fig.~\ref{f:sheath_profiles}(j), however, behaves differently. 
In general, the electron temperature near the cathode increases with bias potential.
However, for bias potentials starting at around \SI{4}{\kilo \volt},
a local maximum develops near the cathode.
As the bias potential continues to increase, this local maximum gets larger and shifts
further away from the cathode (to the left). This is partly due to
the electron density tending to zero at higher bias potentials; 
note how the density is in the denominator in Eq.~\ref{eq:T}.

\begin{figure}[!htb]
	\centering
	\includegraphics[width=\linewidth]{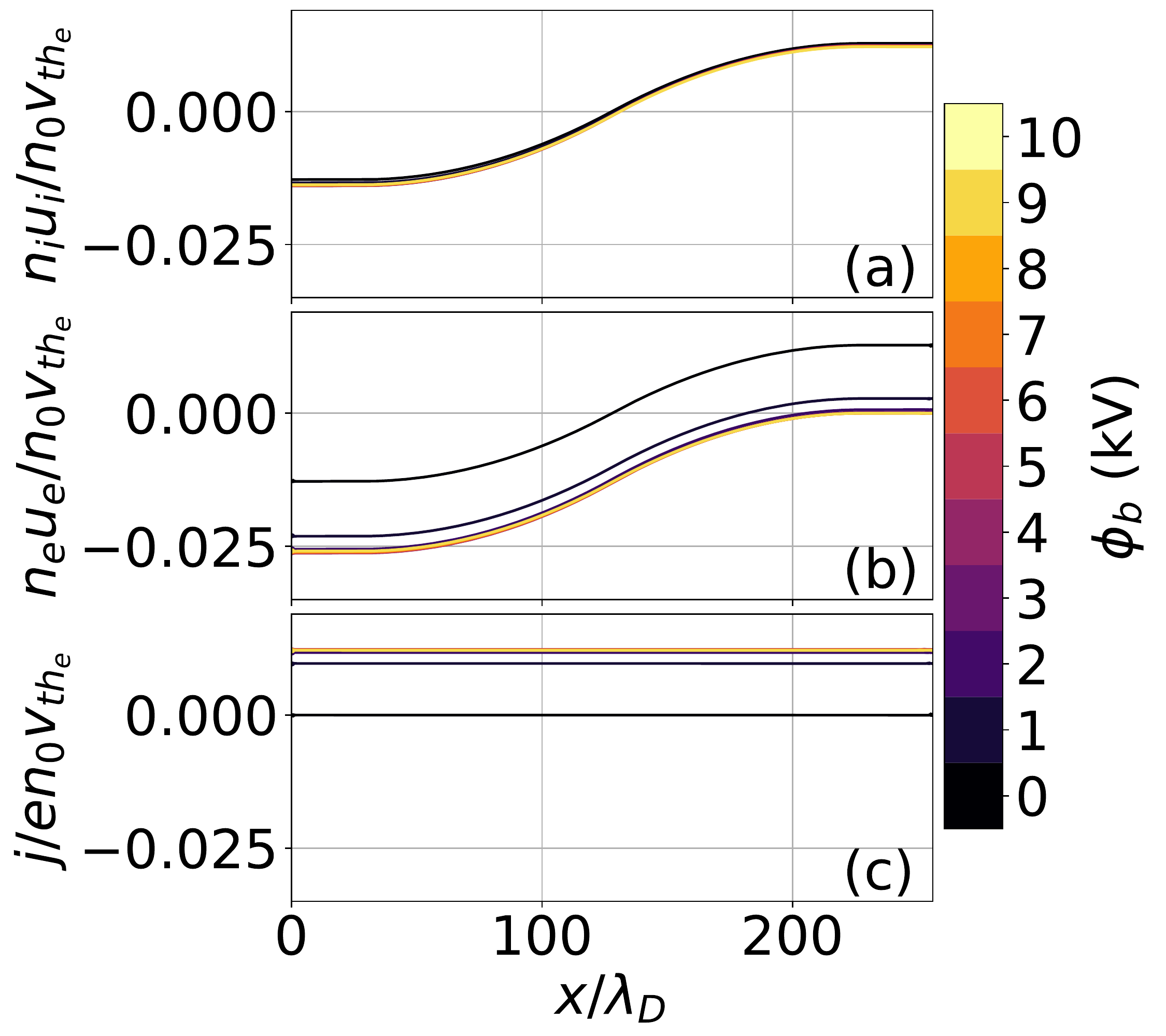}
	\caption{Plots of the normalized ion particle flux (a),
		electron particle flux (b), 
		and current density (c)
		for varying bias potentials. 
		The ion particle flux changes minimally with
		bias potential.
		The electron particle flux and current density
		reach an asymptotic limit as bias potential increases.
	}
	\label{f:current}
\end{figure}

Unlike a classical sheath where there is expected  
to be no current within the plasma, the inclusion of a bias potential
causes a current to develop across the entire domain. Fig.~\ref{f:current}
shows the normalized ion particle flux, electron
particle flux, and current density. All of these quantities use
the electron thermal velocity as one of their normalization
factors to make fair comparisons. 
The ion particle flux, Fig.~\ref{f:current}(a), does not change with 
bias potential retaining its antisymmetry about the center of the domain. 
Therefore, the ion drift velocity in the sheath near the cathode  % Think about this sentence
increases with bias potential because the ion density near the cathode decreases, as shown in Fig.~\ref{f:sheath_profiles}(d).
The electron particle flux, Fig.~\ref{f:current}(b), stays the same 
shape, but is translated in the negative direction % Rethink how I am saying this
as bias potential increases. 
The result is that 
the current density is constant in space
for all bias potentials, as shown in Fig.~\ref{f:current}(c).
Furthermore, as the bias potential increases
the electron particle flux, and therefore the current density,  
converge to asymptotic profiles.
Figs.~\ref{f:current}(a,b) show that, at high bias potentials,
the current is entirely ion driven near the cathode, whereas it 
is predominantly electron driven toward the anode.   % Is this the right place for this sentence?

\begin{figure}[!htb]
	\centering
	\includegraphics[width=\linewidth]{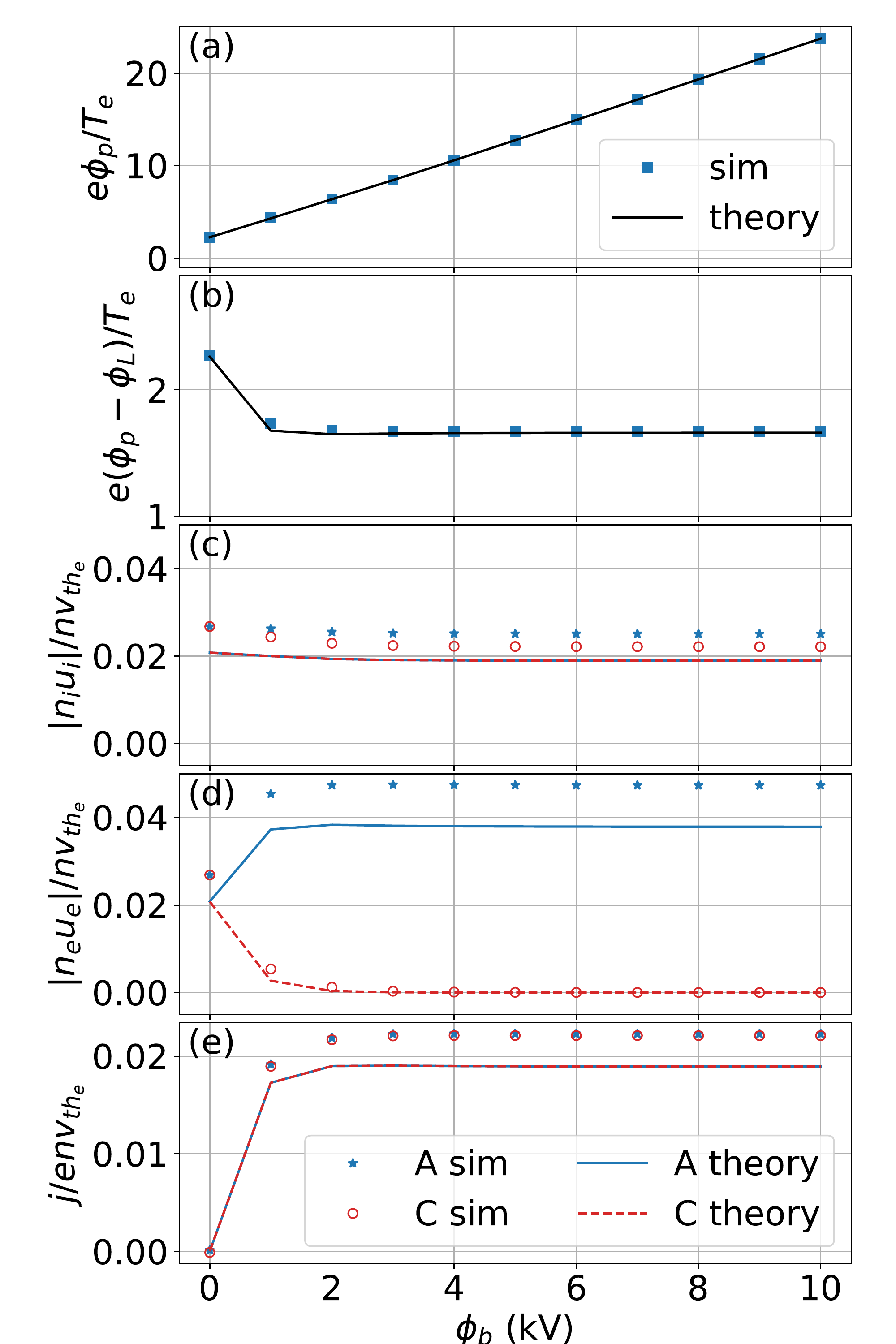}
	\caption{A comparison of the simulaton results
		to theory\cite{stangeby2000plasma} on the effect
		of bias potential.
		Panel (a) shows the normalized plasma potential, defined as
		the potential at the center of the domain. 
		Panel (b) shows the normalized difference between the plasma
		potential and the potential at the anode.
		For Panels (a-b), the squares represent the simulation values
		whereas the solid black line represents the theory.
		Panels (c-e) show the normalized ion particle flux,
		electron particle flux, and current density, respectively,
		at the anode (A) and cathode (C).
		The blue stars and red circles correspond to the simulation
		results at the anode and cathode, respectively.
		The blue solid line and the red dashed line correspond to the 
		theory at the anode and cathode, respectively.
		Note that Panels (c-d) show absolute values to more easily 
		convey the data;
		the particle fluxes at the anode are negative (going into the wall).	
		Unlike the other plots, all normalizations in this plot use the 
		values at the center of the domain.	
	}
	\label{f:compare_stangeby}
\end{figure}

% Perhaps revisit this paragraph
In addition, the electron particle flux tends to 0 near the cathode.
This is a direct result of the chosen boundary condition: the perfectly
absorbing wall. By definition, any electron particle flux
that flows into the cathode (positive $x$ direction) must leave 
the domain. Therefore, directly at the cathode, the minimum possible value
of the electron particle flux is 0. 
Thus, with increasing bias potential,
the current density must saturate to the value of the ion particle flux at the cathode, as 
is shown theoretically in Eq.~\ref{eq:j_sat}.
The current density saturation and how it compared to theory from Eq.~\ref{eq:j_sat}
is examined more closely in the discussion around Fig.~\ref{f:compare_stangeby}. % MADE FIX HERE
Future work will examine what impact wall emissions may have on the current density
and the sheath dynamics.\cite{cagas2020plasma,bradshaw2022}

\begin{figure}[!htb]  
	\centering
	\includegraphics[width=\linewidth]{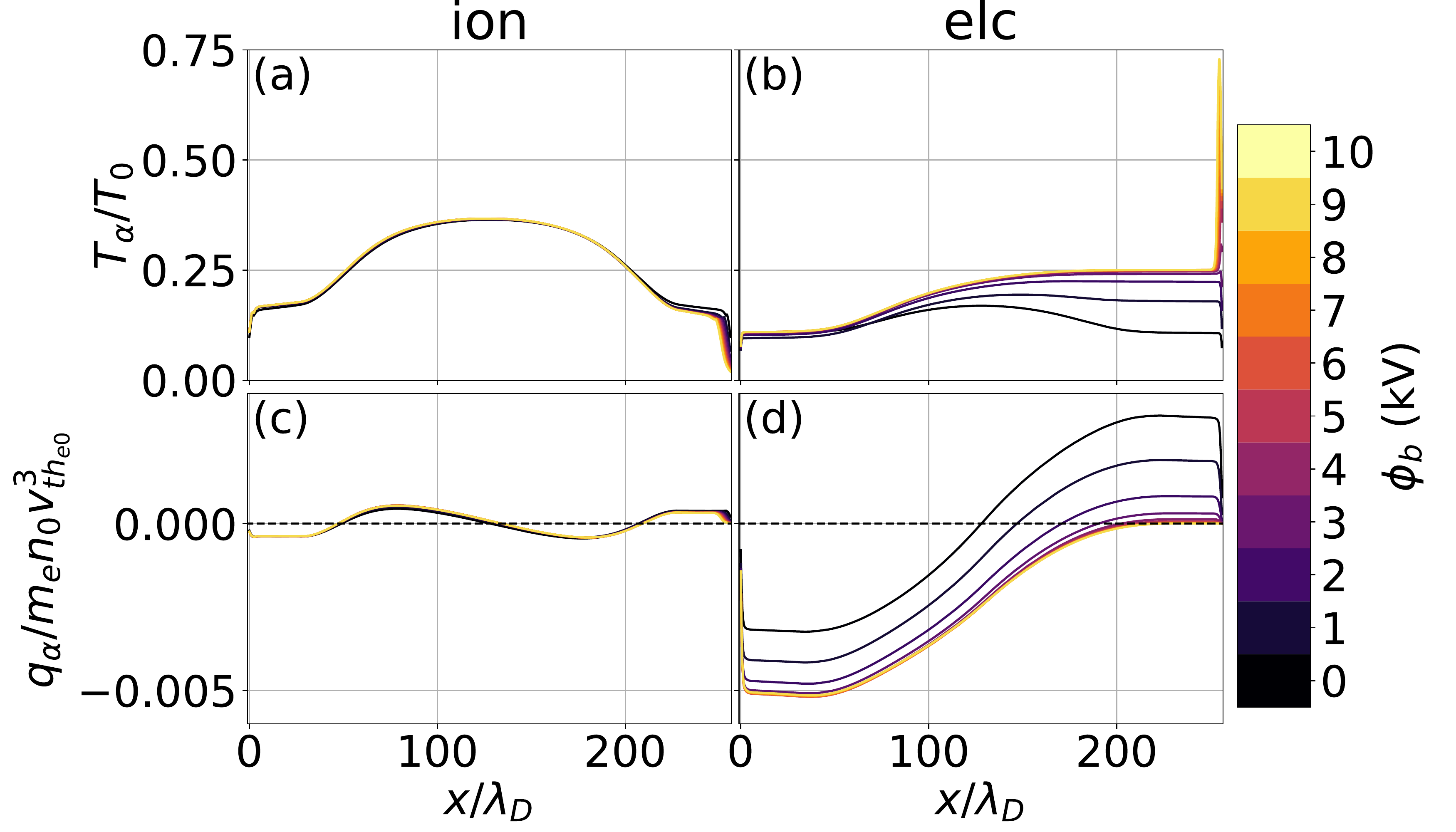}
	\caption{Plots of the normalized temperature (a,b)
		and heat flux (c,d) of 
		the ions (a,c) and electrons (b,d)
		for varying bias potentials.
		The 0 value for the heat fluxes is
		highlighted with a dashed black line.
		The ion properties do not change much with bias 
		potential whereas the electron properties
		change substantially.
	}
	\label{f:temp_q_profiles}
\end{figure}

% Think of a better transition to this paragraph
Fig.~\ref{f:temp_q_profiles} shows that the electrons, unlike the ions, change significantly with bias potential.
Near the anode and in the center of the domain,
the ions remain generally unaffected by the bias potential due to their significantly larger mass than the electrons.
Near the cathode, however, the ion temperature and heat flux decrease with bias potential.

The electron temperature, Fig.~\ref{f:temp_q_profiles}(b), toward the anode
is minimally affected by bias potential. However, this is not the case in the center of the domain
and much less toward the cathode where the
electron temperature increases with increasing bias potential.
Except for near the cathode, the electron temperature profile reaches an asymptotic limit by
about \SI{3}{\kilo \volt}. Near the cathode, the local maximum that develops continues to increase with
increasing bias potential, as also shown in Fig.~\ref{f:sheath_profiles}(j).

The electron heat flux, Fig.~\ref{f:temp_q_profiles}(d), becomes more negative with increasing bias potential.
This matches the trend seen in the electron particle flux, as shown in Fig.~\ref{f:current}(b). 
As expected, for the zero bias potential case, the electron heat flux is antisymmetric about the center.
However, as the bias potential increases, the location where the heat flux is zero (dashed black line)
moves towards the cathode indicating that higher potential increases electron heat flux at the anode.
Eventually, the electron heat flux reaches an asymptotic profile
with the heat flux becoming significant at the anode and negligible at the cathode at larger bias potentials.
The saturation of the electron heat flux is what causes the saturation
of the electron temperature. The spikes in electron temperature near the cathode
are attributed to a significantly decreasing density and a non-Maxwellian distribution. % MADE FIX HERE

It is important to note how the temperatures, Figs.~\ref{f:temp_q_profiles}(a-b),
are significantly lower than what was initialized ($T_0=\SI{2}{\kilo \electronvolt}$).  % MADE FIX HERE
Lower temperatures are expected in simuations of sheaths in the absence of a 
high temperature source that does not 
replenish the energy that is lost to decompressional cooling. 
The impact of a high temperature source will be explored in subsequent work.
Furthermore, the simulation results shown thus far are in only one velocity dimension.
This inherently assumes that the plasma temperature is isotropic, which is not the case for 
a plasma sheath.\cite{cagas2017continuum}
The inclusion of more velocity dimensions, as discussed in Sec.~\ref{s:1x2v},
allows for better thermalization through collisions with the hotter
perpendicular temperature.

Fig.~\ref{f:compare_stangeby} shows how well
certain values from simulation
compare with the theory from Sec.~\ref{s:theory}.
The theory, denoted by the lines, is calculated using the
densities and temperatures at the center of the domain.
The axis normalizations also use the center values.

Fig.~\ref{f:compare_stangeby}(a) shows the comparison of the 
plasma potential, defined as the electric potential in the center 
of the domain, between simulation (blue squares)
and theory (solid black line) based on Eq.~\ref{eq:phi_p}.
The simulation is well predicted by the theory.
Since the ground potential
is defined to be at the cathode, the plasma potential is equivalent to the 
potential drop between the center of the domain and cathode.
Fig.~\ref{f:compare_stangeby}(b) shows how the potential drop at the anode
changes with bias potential.
In corroboration with Fig.~\ref{f:sheath_profiles}(a), the 
potential difference at the anode remains relatively constant with bias potential.
The simulation results are compared to the theoretical difference,
which is calculated as $\tilde{\phi}_p-\tilde{\phi}_b$, or the difference
between Eq.~\ref{eq:phi_p} and $\tilde{\phi}_b$.
Again, the simulation results agree well with theory.

Fig.~\ref{f:compare_stangeby}(c) compares the 
ion particle flux at the electrodes with 
theory, Eqs.~\ref{eq:niui_lw} and \ref{eq:niui_rw}, respectively. 
The flux at the anode changes slightly with bias potential,
whereas the flux at the cathode remains constant;
both are higher than predicted by theory.
Fig.~\ref{f:compare_stangeby}(d) compares the 
electron particle flux at the anode and cathode with 
theory, Eqs.~\ref{eq:neue_lw} and \ref{eq:neue_rw}, respectively. 
The electrons follow the general trend of the theory, but saturate
to a higher value than predicted.
The electron particle flux at the cathode
decreases with bias potential and asymptotes to 0, matching theory.
Fig.~\ref{f:compare_stangeby}(e) compares the 
current density at the electrodes with theory, Eq.~\ref{eq:j}.
As also shown in Fig.~\ref{f:current}(c), the current density at the
electrodes are the same, with the points lying on top of each other.
For the classical sheath case, the theoretical result is exactly met with 
there being no current in the plasma. 
For every other case, the current density is larger than predicted by theory.

The potential differences match well with theory.
However, there are differences between theory and simulation for
the particle fluxes, and therefore the current density.
The theory assumes an isothermal fluid plasma with Boltzmann distributed electrons.\cite{stangeby2000plasma}
As shown in Figs.~\ref{f:sheath_profiles}(g-j) and \ref{f:temp_q_profiles}(a-b),
the temperature is not isothermal. 
Therefore the Bohm speed used in Eq.~\ref{eq:niui_lw} and \ref{eq:niui_rw}
is expected to have $\gamma_e=\gamma_i=3$ (for a 1D plasma) 
instead of 1, which would increase the current.\cite{li2022bohm} 
Furthermore, the simulation model is fully kinetic and evolves the entire
distribution function. 
Thus, there are additional kinetic physics that are not captured within the theory, such as the impact of heat flux. 
Therefore, the simplified theory underpredicts the actual particle fluxes and current density.

Fig.~\ref{f:wall_fluxes_SI} shows the values of the normalized
density, ion drift velocity, temperature,
heat flux, current, and sheath length at the electrodes
as functions of the bias potential.
These are reported to provide insight into the conditions the electrodes
face which can inform future design choices.
In line with the findings of the anode profiles in Fig.~\ref{f:sheath_profiles},
the density, drift velocity, and temperature for
the ions and electrons, Figs.~\ref{f:wall_fluxes_SI}(a,c,e), change minimally with bias potential.  
While the ion heat flux negligibly changes with bias potential,
the electron heat flux slightly increases with bias potential, reaching an 
asymptotic value, as shown in Fig.~\ref{f:wall_fluxes_SI}(g).

The story is different at the cathode for Figs.~\ref{f:wall_fluxes_SI}(b,d,f,h). 
The ion density and temperature both decrease with increasing bias potential.
The ion drift velocity increases significantly with bias potential, which should be 
expected based on ion particle flux remaining constant but the ion density decreasing.
The ion heat flux tends to zero at the cathode.
The electron properties at the cathode show much larger changes.
The electron density and heat flux tend to zero at the cathode
with increasing bias potential. The electron temperature at the cathode, however,
increases with bias potential; it increases quickly at first, but then the rate slows down
suggesting the possibility of an asymptotic limit at even higher bias potentials.
Both the ion and electron heat fluxes decrease significantly with bias potential with
both of them approaching 0 by approximately \SI{3}{\kilo \volt}.

Thus, based on Figs.~\ref{f:compare_stangeby}(c-d) and
\ref{f:wall_fluxes_SI}(g-h), the particle and heat fluxes
that the anode and cathode face are different. 
For high bias potentials, both electrodes face the same ion particle flux.
However, the anode faces approximately twice hte 
amount of electron particle flux while the cathod
faces almost no electron particle flux.
Additionally, at the anode, the ion heat flux remains approximately
constant with bias potential while the electron heat flux increases significantly.
However, the ion and electron heat fluxes both tend to zero
at the cathode for high bias potentials.
Therefore, the material choice for the anode
needs to account for larger particle and heat fluxes,
whereas this is less of a design concern for the cathode.

Figs.~\ref{f:wall_fluxes_SI}(i-j) show the current in \si{\kilo \ampere}
at the anode and cathode, respectively. The current is calculated assuming
it is perfectly uniform across a cylindrical Z-pinch.
In other words, the current is $I = j \pi a^2$, where $a$ is the pinch radius.
For the FuZE parameter regime,\cite{zhang2019sustained} the pinch radius is \SI{3}{\milli \meter}.
As expected from Fig.~\ref{f:current}(c), the currents at both electrodes
are identical. FuZE reports a current of about \SI{200}{\kilo \ampere},\cite{zhang2019sustained} whereas 
the simulation current asymptotes to about \SI{115}{\kilo \ampere}. The differences
are largely explained by the lower temperature in the domain and assumptions made for the radial current profile.
A similar calculation based purely on the simplified theory of Eq.~\ref{eq:j_sat} using the initial parameters
yields a current of \SI{154}{\kilo \ampere}. If instead we use the
simulation density and temperature at the center of the domain, we obtain a current of \SI{97}{\kilo \ampere},
which is lower than the simulation current as is also shown in Fig.~\ref{f:compare_stangeby}(e);
the theory underpredicts the current that develops.
Therefore, it is anticipated that in a case with less decompressional cooling in the domain such that
the steady state temperature is closer to the initialized temperature, the current output will
be higher than the predicted \SI{154}{\kilo \ampere}.

Figs.~\ref{f:wall_fluxes_SI}(k-l) show how the sheath lengths change with bias potential.
The sheath lengths are calculated in three ways: 1\% fractional charge density, 
10\% fractional charge density, and Bohm speed.
The fractional charge density is defined as $|n_i-n_e|/(n_i+n_e)$.
While more sophisticated theory has been developed for the Bohm speed accounting 
for transport and anisotropies,\cite{li2022bohm,li2022transport} 
a simplified Bohm speed is used here: $u_B=\sqrt{\gamma (T_i+T_e) / m_i}$, where $\gamma$
is the ratio of specific heats. For one velocity dimension (or degree of freedom), $\gamma$ is 3.
The sheath length is determined based on the location where
the ratio of the ion velocity to the Bohm speed, $u_i/u_B$, is 1
or where the the fractional charge density is either 1\% or 10\%.
The results for the anode show that the sheath length does not change
significantly with bias potential. For the cathode,
the sheath length increases with bias potential as expected based on Figs.~\ref{f:sheath_profiles}(d,f). 
This is because the electrons, 
being much less massive, are impacted more heavily by the larger bias potentials
creating a larger region of positive charge.

\subsection{Comparison with 1X-2V} \label{s:1x2v}

Due to the large computational expense, one 1X-2V simulation is performed with a \SI{5}{\kilo \volt}
bias potential to understand how the additional velocity dimension
impacts the solution. 
The results presented here are at $t=20000\omega_{pe}$, by which point the plasma has 
reached a quasi-steady state. 

\begin{figure}[!htb] 
	\centering
	\includegraphics[width=\linewidth]{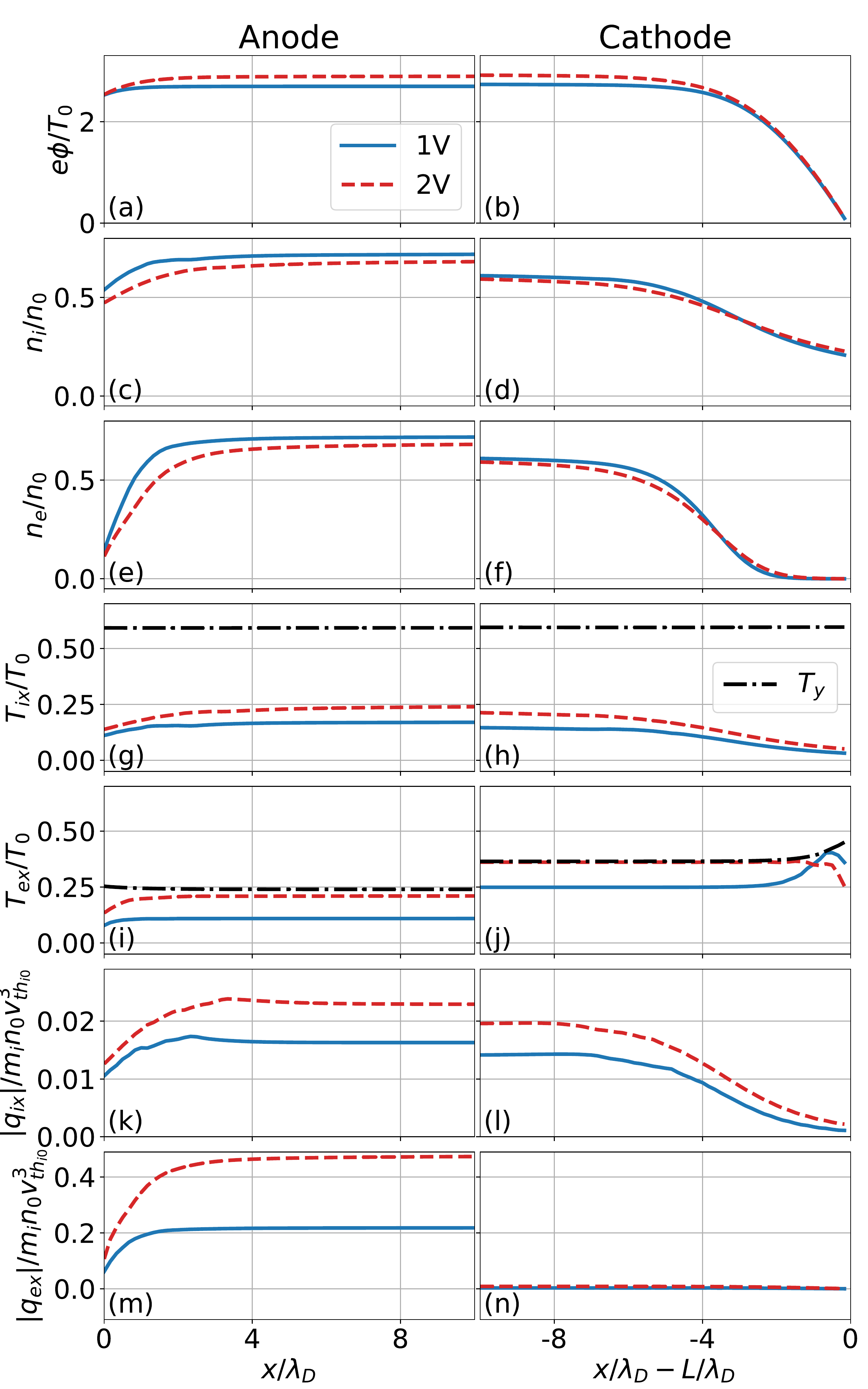}
	\caption{Comparison of normalized sheath profiles
		for the electric potential (a,b), 		
		densities (c-f), 
		temperatures (g-j), and
		heat fluxes (k-n)
		between a 1X-1V (solid blue line) and 1X-2V (dashed red line)
		simulation for $\phi_b=\SI{5}{\kilo \volt}$.
		The black dash-dot line in Panels (e-h) represents the $y$, or perpendicular, direction
		temperature from the 1X-2V simulations.
		In general, the differences between the profiles near the cathode are minimal,
		whereas they are more substantial at the anode.
	}
	\label{f:sheath_comp_5}
\end{figure}

Fig.~\ref{f:sheath_comp_5} shows a comparison of the 1X-1V results
to the 1X-2V results for sheath profiles near the electrodes
of the electric potential, densities, temperatures, and heat fluxes.
This plot is directly analagous to Fig.~\ref{f:sheath_profiles}.
Figs. \ref{f:sheath_comp_5}(a-b) show the electric potential near the electrodes. 
For about 1 Debye length near the anode and about 4 Debye lengths near the cathode,
the electric potential for the 1X-1V and 1X-2V cases are approximately the same.
However, the electric potential towards the center of the domain, or the plasma potential,
of the 1X-2V case is larger than that of the 1X-1V case.  
Therefore, the potential drop is larger at both electrodes, which results
in sharper sheath phenomena. 
Relatively, the difference in the potential drop
is larger at the anode than at the cathode. 
In other words,
the potential difference between the center of the domain and cathode is already large for the 1X-1V case
and only increases in the 1X-2V case by 9.4\%.
The potential difference between the center of the domain and the anode is comparatively small for the 1X-1V case
and increases in the 1X-2V case by 71.9\%.
Therefore, the anode sheath is affected more prominently than the cathode sheath.

Following the pattern of the electric potential,
the ion and electron density near the cathode, Figs.~\ref{f:sheath_comp_5}(d,f), 
only show a minimal decrease while the densities near the anode, Figs.~\ref{f:sheath_comp_5}(c,e),
show a more notable decrease.

The temperatures in the domain are generally larger, as seen in Figs.~\ref{f:sheath_comp_5}(g-j). 
In addition, due to considering kinetic physics, the particle distribution, unlike in
the fluid approximation, is not required to be isotropically Maxwellian.
Therefore, temperature, as defined by Eq.~\ref{eq:T}, can be different
in each direction. 
Here, we define parallel ($x$) as in the direction of the flow.
Therefore the $y$ direction is the perpendicular direction. % MADE FIX HERE
In the parallel direction, there exists a heat sink with the perfectly absorbing wall
boundary condition.
In the perpendicular direction, however, there is no such heat sink.
Therefore, an anisotropy develops in the temperature. 
The dash-dot black line depicts the 
temperature in the $y$ or perpendicular direction. 
Unlike the parallel temperature,
the perpendicular temperature increases in the sheath;
this increase is much larger at the cathode than at the anode.
This is due to the gradient of the parallel heat flux
of the perpendicular degrees of freedom.\cite{zhang2022resolving}

The ion heat fluxes are larger in the 1X-2V simulation, as shown in Figs.~\ref{f:sheath_comp_5}(k-l).
The electron heat flux, Figs.~\ref{f:sheath_comp_5}(m-n), is only larger near the anode;
near the cathode, it remains close to zero as with the 1X-1V case.

\begin{figure}[!htb]  
	\centering
	\includegraphics[width=.8\linewidth]{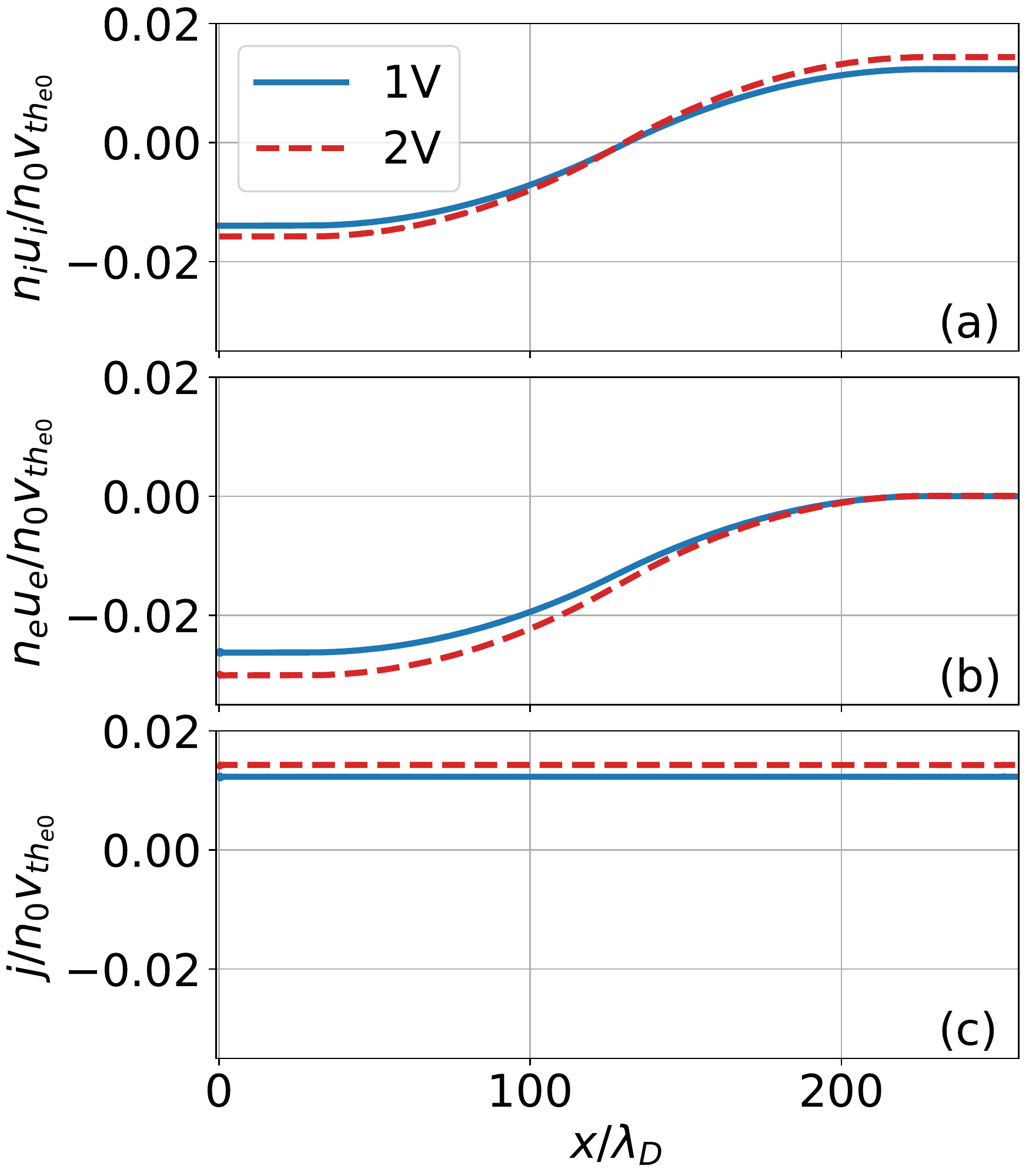} 
	\caption{Comparisons of	the normalized ion particle flux (a),
		electron particle flux (b), and current density (c) between
		a 1X-1V (solid blue line) and 1X-2V (dashed red line) simulation
		with $\phi_b=\SI{5}{\kilo \volt}$.
	}
	\label{f:current_comp_5}
\end{figure}

Fig.~\ref{f:current_comp_5} shows the comparison of
the normalized particle flux and current between
the 1X-1V and 1X-2V simulations.
This figure is directly analogous to Fig.~\ref{f:current}.
The ion particle flux, Fig.~\ref{f:current_comp_5}(a), is 13.1\% larger in magnitude
in the 1X-2V case. This is due to the higher general temperature
in the domain. The electron particle flux, Fig.~\ref{f:current_comp_5}(b),
still goes to zero at the right wall, but is 14.4\% larger in magnitude at the anode.
The result is that the current density, Fig.~\ref{f:current_comp_5}(c), is 15.9\% larger in the 
1X-2V case. The current density is still spatially constant.
If a similar calculation is done as that in Figs.~\ref{f:wall_fluxes_SI}(i-j) 
to obtain an estimated current assuming a perfectly uniform Z-pinch, we find that
the 1X-2V case yields a current of \SI{134}{\kilo \ampere}.

% Make aspect ratio of this plot a bit better. Doesn't look as nice.
\begin{figure}[!htb] 
	\centering
	\includegraphics[width=\linewidth]{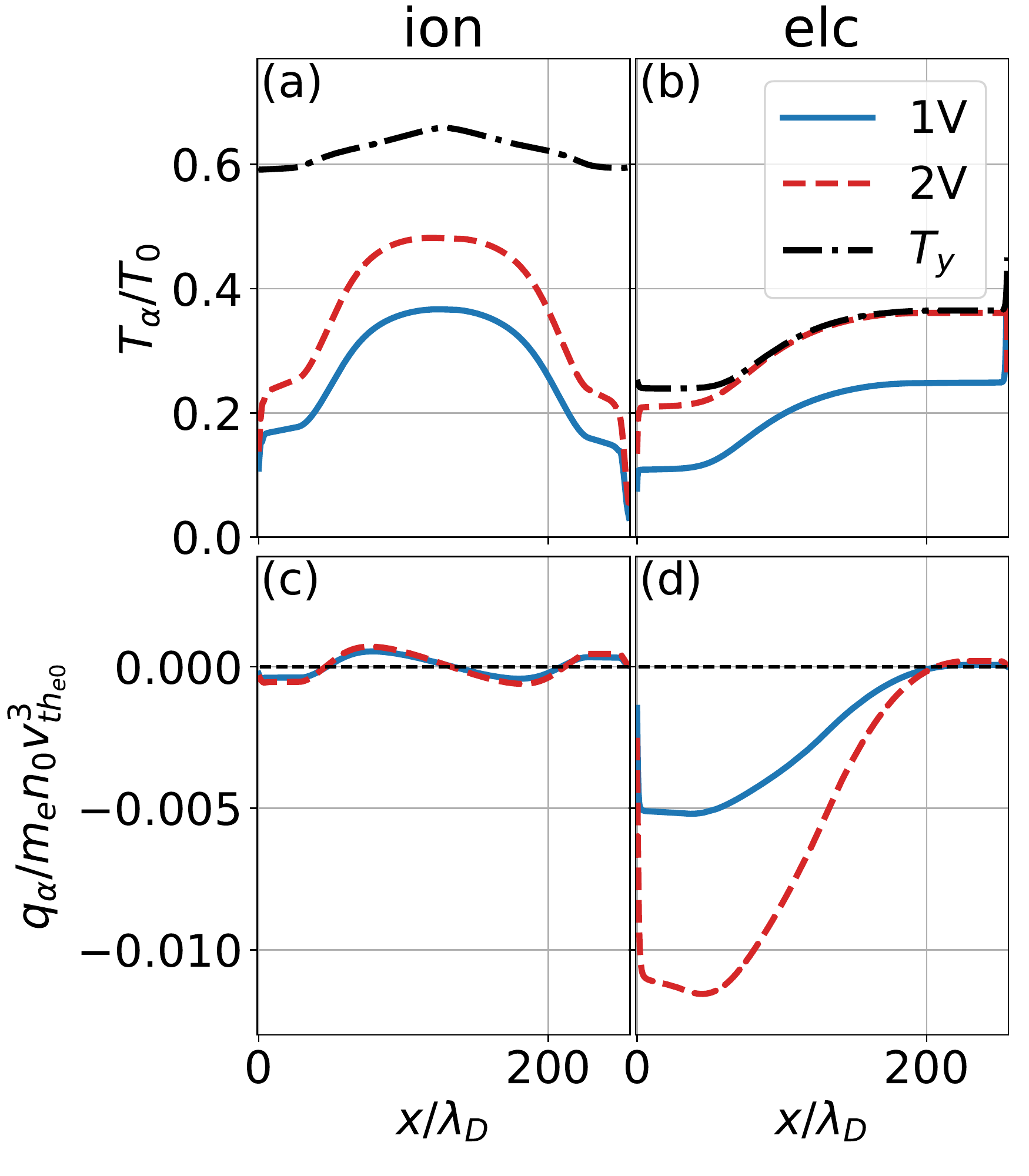} 
	\caption{Comparisons of normalized temperature (a,b)
		and heat flux (c,d) for ions (a,c) and electrons (b,d)
		between the 1X-1V (solid blue line) and 1X-2V (dashed red line)
		simulations with $\phi_b=\SI{5}{\kilo \volt}$.
		The black dash-dot line represents the temperature in the $y$ (perpendicular)
		direction.
	}
	\label{f:comp_qT}
\end{figure}

Fig.~\ref{f:comp_qT} compares how the temperatures and heat fluxes
vary between the 1X-1V and 1X-2V simulations.
The inclusion of the second velocity dimension allows
for an anisotropy to develop in the sheath region for the temperature.
Note that there is still cooling in 1X-2V, 
but since decompressional cooling primarily impacts the parallel temperature, 
the presence of the perpendicular temperature with presheath 
collisions permits higher temperatures in the domain (compared to 1X-1V) late in time.
This is the primary reason why the plasma potential and the ion particle flux are larger
for the 1X-2V simulations.
The electron temperature shows anisotropy only near the electrodes, with the
parallel and perpendicular temperatures being the same
in the center of the domain. 
However, an anisotropic equilibrium develops for the ion temperature due to
the energy loss in the parallel direction being larger than that of the perpendicular direction 
and the collisional isotropization.   % MADE FIX HERE

The ion heat flux does not change much with the inclusion of the
second velocity dimension compared to how much the
electron heat flux changes for the 1X-2V case. 
Near the cathode, the electron heat flux remains close to zero.
However, it increases sharply towards the anode and has a much larger drop
at the anode. This suggests that the 1X-1V simulations underpredicts the electron heat
flux for the entire domain apart from near the cathode.
Therefore, even more care needs to be taken in material choice at the anode
due to the higher fluxes.

\section{Summary and Conclusions} \label{s:conclusions}

Continuum kinetic simulations of a proton-electron plasma are performed solving the Boltzmann-Poisson system
to model the sheath formation of a plasma bounded between two walls, or electrodes, with 
an electric potential bias between them. The parameter regime is chosen to be 
relevant to Z-pinch fusion experiments.

1X-1V simulations are performed with bias potentials
ranging from 0 to \SI{10}{\kilo \volt}. The results suggest that
the sheath formation at the left wall, or anode, does not change greatly with bias potential.
The right wall, or cathode, however, exhibits a much steeper potential drop, resulting in greater
differences in sheath formation. 
Near the cathode, the ion and electron densities, ion temperature,
and ion and electron heat fluxes decrease with bias potential.  
The electron temperature, however, increases with bias potential.

The ion particle flux does not change significantly within the entire domain with varying bias potential.
The electron particle flux becomes more negative with increasing bias potential, reaching
an asymptotic limit when the particle flux at the cathode reaches zero.
For all cases, the current density is spatially constant and increases with bias potential.
Because the ion particle flux does not change and the electron particle flux reaches an asymptotic limit,
the current density also reaches a limit with increasing bias potential.
The current is predominantly carried by the electrons at the anode
and by the ions at the cathode.

The 1X-1V simulations match well with theory for
the plasma potential. The simulation trends well with the 
theory for the particle fluxes. However, the theory underpredicts
the simulation results for current and fluxes. % MADE FIX HERE
The differences are explained by additional kinetic physics and fewer assumptions considered
in the simulations.

Particle and heat fluxes for the ions and electrons at both electrodes are provided.
Assuming the current is carried uniformly in the Z-pinch with FuZE parameters, the current
from the simulations asymptote to \SI{115}{\kilo \ampere} with increasing bias potential, as compared
to the \SI{200}{\kilo \ampere} measured in FuZE. 

At the anode, the ion particle and heat fluxes remain relatively constant with 
bias potential; the electron particle and heat fluxes increases by a factor of approximately
2 at high bias potentials. 
At the cathode, however, the electron particle flux, electron heat flux, and ion heat flux
tend to zero at higher bias potentials, whereas the ion particle flux remains constant.
Therefore for high bias potentials,
the material choice for the anode needs to consider higher fluxes from the ions and electrons;
for the cathode, only the ion particle flux needs to be considered.

A 1X-2V simulation is run for the \SI{5}{\kilo \volt} bias potential case.
The results show a temperature anisotropy develop in the sheath region.
In general, the sheath effects are more stark than those from the 1X-1V simulations.
The increase is attributed to the second velocity dimension causing less overall
cooling in the system due to collisions involving the perpendicular temperature.
The increased overall temperature of the plasma also contributes to
a 15.9\% increase in the current.
Simulations performed with higher velocity dimensions produce reduced cooling and larger currents,
which are closer to the experimental values.

Future work will leverage recent developments that may permit 
computationally-efficient calculations using higher velocity dimensions.  Additionally, the impact 
of physical wall materials on sheaths will be studied with biased potentials in subsequent work.

\begin{acknowledgments}
	This work was supported by the Department of Energy ARPA-E BETHE program under award number DE-AR0001263.  
	
	J. Juno acknowledges support by the U.S. Department of Energy under Contract No. DE-AC02-09CH1146 via an LDRD grant.
	
	The authors thank Petr Cagas for useful discussions on sheaths.
	
	The authors acknowledge Advanced Research Computing at Virginia Tech for providing computational resources and technical support that have contributed to the results reported within this paper. URL: \url{http://www.arc.vt.edu}

\end{acknowledgments}

\section*{Data Availability Statement}
Readers may reproduce our results and also use Gkeyll for their applications. The code and input files used here are available online. 
Full installation instructions for Gkeyll are provided on the Gkeyll website.\cite{gkylDocs} The code can be installed on Unix-like 
operating systems (including Mac OS and Windows using the Windows Subsystem for Linux) either by installing the pre-built binaries using 
the conda package manager (\url{https://www.anaconda.com}) or building the code via sources. The input files used here are under version 
control and can be obtained from the repository at \url{https://github.com/ammarhakim/gkyl-paper-inp/tree/master/2022_PoP_BiasedWallSheaths}.

\appendix

\section{1D Initial Sheath Approximation}
\label{a:Robertson_solver}

The initial conditions used in the simulations (see Fig.~\ref{f:robertson_profiles})
are calculated based on theory from Ref.~\onlinecite{robertson2013sheaths}.
With some a priori predictions of the electric potential drop near the wall
and the ion particle flux at the wall, theoretical profiles can be obtained
of the ion density, electron density, ion velocity, and electric potential.

The system of ordinary differential equations (ODEs) that 
describe the steady state sheath behavior are\cite{robertson2013sheaths}
\begin{align}
	\frac{d\tilde{\phi}}{d\tilde{x}} &= - \tilde{E} 
	\label{eq:Rob_ODE_phi} \\
	\frac{d \tilde{E}}{d \tilde{x}} &= \frac{\tilde{R}\tilde{x}}{\tilde{u}_i} - \exp \tilde{\phi} 
	\label{eq:Rob_ODE_E} \\
	\frac{d \tilde{u}_i}{d \tilde{x}} &= \frac{\tilde{E}}{\tilde{u}_i} - \frac{\tilde{u}_i}{\tilde{x}},
	\label{eq:Rob_ODE_ui}
\end{align}
where $R$ is an ionization source term and the tildes denote normalized variables.
These ODEs must be solved numerically. 
The Python package \verb|odeint|\footnote{https://docs.scipy.org/doc/scipy/reference/generated/scipy.integrate.odeint.html}  % Maybe editors know a better way of doing this?
is used for the numerical integration
and uses a combination of the nonstiff Adams method and the stiff backwards difference method.\cite{petzold1983automatic}

Consider a domain that spans from 0 to $\tilde{L}=L/\lambda_D$ where $L$ is the domain length.
At $\tilde{x}=0$, the normalized ion and electron densities are 1
and the normalized electric potential, electric field, and ion velocity are 0.
The starting point for the integration, however, cannot be $\tilde{x}=0$ due to division by
zero (see Eq.~\ref{eq:Rob_ODE_ui}). Instead, the integration begins at $\tilde{x}=\Delta \tilde{x}$
where $\Delta\tilde{x}= \tilde{L}/N$ where $N$ is the number of points considered in the domain.
Thus, the initial conditions for the electric potential, electric field,
and velocity are $-\tilde{R}^2 \Delta \tilde{x}^2$, $2 \tilde{R}\Delta \tilde{x}$,
and $\tilde{R}\Delta \tilde{x}$, respectively.\cite{robertson2013sheaths}

Ref.~\onlinecite{robertson2013sheaths} suggests integrating in $\tilde{x}$ until
the potential drop at the wall reaches a pre-defined value. 
The a priori value can be calculated as
the difference between the plasma potential, Eq.~\ref{eq:phi_p},
and the potential at the wall.
However, there is a second constraint at the wall that must also be met;
based on the steady state continuity equation, the ion particle flux at the wall 
must be $\tilde{n}_i \tilde{u}_i|_{w} = \tilde{R}\tilde{L}$.\cite{robertson2013sheaths}
If we just solve the system of ODEs until the potential drop is reached, the ion particle flux
constraint is not necessarily met. This issue is resolved by iterating on $\tilde{L}$ until both
the potential and particle flux constraints are met.

The inputs of the developed algorithm are the potential drop
and the ion particle flux at the wall. 
Note that this algorithm assumes that the plasma potential (the potential in the center of the domain)
is the ground potential. Therefore, the input potential drop should be negative.
Parts of this algorithm are adapted from a code in Sec.~5.1.3 of Ref.~\onlinecite{cagas2018continuum}.
An arbitrarily small $\tilde{L}$ is chosen
as the initial domain length. Therefore, the ionization source term must be
$\tilde{R}=\tilde{n}_i \tilde{u}_i|_{w}/\tilde{L}$. 
The ODEs are numerically integrated across the entire domain using \verb|odeint|.
Then the value at the wall is tested for the electric potential constraint; 
note that the ion particle flux constraint is automatically satisfied through the definition
of $\tilde{R}$. The domain length is iterated on systematically
to reach the correct solution.
Additionally, define a switch, $r$, to note when the electric potential
at the wall first becomes less than the input potential. 
Due to the assumed monotonicity
of the potential profile, this is equivalent to the tested domain length being larger
than the final converged domain length.
Alg.~\ref{alg:rob} describes how we iteratively converge to the solution.

% Is this best practice for this? 
\begin{algorithm}[!htb]
	\caption{Algorithm to solve Eqs.~\ref{eq:Rob_ODE_phi}-\ref{eq:Rob_ODE_ui} iteratively
		by changing domain length\label{alg:rob} }
	\begin{algorithmic}[1]
		\Require $\Delta \tilde{\phi}$ and $\tilde{n}_i\tilde{u}_i|_w$   % This is the input to the function
		\Ensure Initial condition profiles for $\tilde{n}_i$, $\tilde{n}_e$, $\tilde{u}_i$, and $\tilde{\phi}$		
		\State Set initial arbitrarily small $\tilde{L}_0$
		\For{$i \gets 0$ to max\_iter}
		\State Build $\tilde{x}$ grid
		\State $\tilde{R} \gets \tilde{n}_i\tilde{u}_i|_w/\tilde{L}_i$
		\State Set ODE initial conditions
		\State Solve ODEs using \verb|odeint|
		\If{$\Delta \tilde{\phi} - TOL \leq \tilde{\phi}_w \leq \Delta \tilde{\phi} + TOL$}
		\State break
		\ElsIf{$\tilde{\phi}_w > \Delta \tilde{\phi}$ and $r = 0$}
		\State $\tilde{L}_{i+1} \gets 2 \tilde{L}_i$
		\ElsIf{$\tilde{\phi}_w > \Delta \tilde{\phi}$ and $r = 1$}
		\State $\tilde{L}_{i+1} \gets \tilde{L} + \Delta \tilde{L}_i/2$
		\ElsIf{$\tilde{\phi}_w < \Delta \tilde{\phi}$}
		\State $r \gets 1$
		\State $\tilde{L}_{i+1} \gets \tilde{L} - \Delta \tilde{L}_i/2$
		\EndIf
		\State $\Delta \tilde{L} \gets |\tilde{L}_{i+1}-\tilde{L}_i|$
		\EndFor
	\end{algorithmic}
\end{algorithm}

There are two known failure modes for this algorithm.
It is possible that the solution converges to some $\tilde{L}$ at which the
electric potential is not within the tolerance ($TOL$) of the input potential. In this case, 
if $\Delta\tilde{L}$ is less than the tolerance, then the solution is assumed to have converged
to the best possible solution.
The other failure mechanism is that the solution can diverge and the 
electric potential constraint will never be satisfied; therefore, no solution can be found.
These failure modes arise based on the inputs of the ion particle flux and electric potential
at the wall. For ion particle fluxes that are too low (generally lower than 0.51), the solution diverges.
Additionally, the electric potential at the wall must be negative.

The profiles in Fig.~\ref{f:robertson_profiles} are found by using 
Alg.~\ref{alg:rob}. The ion particle flux input is set
as $\tilde{n}_i\tilde{u}_i|_w=0.55$. The potential input is set
based on the definitions in Sec.~\ref{s:theory}; the potential drops used for
the anode and cathode are $\Delta \tilde{\phi}|_A=\tilde{\phi}_b - \tilde{\phi}_p$ and
$\Delta \tilde{\phi}|_C= -\tilde{\phi}_p$, respectively.
To calculate the plasma potential, additional inputs of the ion mass, ion temperature,
and electron temperature are needed. For these cases, we use the proton mass and 
$T_i=T_e=\SI{2}{\kilo \electronvolt}$.

\bibliography{reference}

%aipnum4-2.bst 2019-01-14 (MD) hand-edited version of apsrev4-1.bst
\begin{thebibliography}{39}%
\makeatletter
\providecommand \@ifxundefined [1]{%
 \@ifx{#1\undefined}
}%
\providecommand \@ifnum [1]{%
 \ifnum #1\expandafter \@firstoftwo
 \else \expandafter \@secondoftwo
 \fi
}%
\providecommand \@ifx [1]{%
 \ifx #1\expandafter \@firstoftwo
 \else \expandafter \@secondoftwo
 \fi
}%
\providecommand \natexlab [1]{#1}%
\providecommand \enquote  [1]{``#1''}%
\providecommand \bibnamefont  [1]{#1}%
\providecommand \bibfnamefont [1]{#1}%
\providecommand \citenamefont [1]{#1}%
\providecommand \href@noop [0]{\@secondoftwo}%
\providecommand \href [0]{\begingroup \@sanitize@url \@href}%
\providecommand \@href[1]{\@@startlink{#1}\@@href}%
\providecommand \@@href[1]{\endgroup#1\@@endlink}%
\providecommand \@sanitize@url [0]{\catcode `\\12\catcode `\$12\catcode
  `\&12\catcode `\#12\catcode `\^12\catcode `\_12\catcode `\%12\relax}%
\providecommand \@@startlink[1]{}%
\providecommand \@@endlink[0]{}%
\providecommand \url  [0]{\begingroup\@sanitize@url \@url }%
\providecommand \@url [1]{\endgroup\@href {#1}{\urlprefix }}%
\providecommand \urlprefix  [0]{URL }%
\providecommand \Eprint [0]{\href }%
\providecommand \doibase [0]{https://doi.org/}%
\providecommand \selectlanguage [0]{\@gobble}%
\providecommand \bibinfo  [0]{\@secondoftwo}%
\providecommand \bibfield  [0]{\@secondoftwo}%
\providecommand \translation [1]{[#1]}%
\providecommand \BibitemOpen [0]{}%
\providecommand \bibitemStop [0]{}%
\providecommand \bibitemNoStop [0]{.\EOS\space}%
\providecommand \EOS [0]{\spacefactor3000\relax}%
\providecommand \BibitemShut  [1]{\csname bibitem#1\endcsname}%
\let\auto@bib@innerbib\@empty
%</preamble>
\bibitem [{\citenamefont {Stangeby}(2000)}]{stangeby2000plasma}%
  \BibitemOpen
  \bibfield  {author} {\bibinfo {author} {\bibfnamefont {P.~C.}\ \bibnamefont
  {Stangeby}},\ }\href@noop {} {\emph {\bibinfo {title} {The plasma boundary of
  magnetic fusion devices}}},\ Vol.\ \bibinfo {volume} {224}\ (\bibinfo
  {publisher} {Institute of Physics Pub. Philadelphia, Pennsylvania},\ \bibinfo
  {year} {2000})\BibitemShut {NoStop}%
\bibitem [{\citenamefont {Lieberman}\ and\ \citenamefont
  {Lichtenberg}(2005)}]{lieberman2005principles}%
  \BibitemOpen
  \bibfield  {author} {\bibinfo {author} {\bibfnamefont {M.~A.}\ \bibnamefont
  {Lieberman}}\ and\ \bibinfo {author} {\bibfnamefont {A.~J.}\ \bibnamefont
  {Lichtenberg}},\ }\href@noop {} {\emph {\bibinfo {title} {Principles of
  plasma discharges and materials processing}}}\ (\bibinfo  {publisher} {John
  Wiley \& Sons},\ \bibinfo {year} {2005})\BibitemShut {NoStop}%
\bibitem [{\citenamefont {Baalrud}\ \emph {et~al.}(2020)\citenamefont
  {Baalrud}, \citenamefont {Scheiner}, \citenamefont {Yee}, \citenamefont
  {Hopkins},\ and\ \citenamefont {Barnat}}]{baalrud2020interaction}%
  \BibitemOpen
  \bibfield  {author} {\bibinfo {author} {\bibfnamefont {S.~D.}\ \bibnamefont
  {Baalrud}}, \bibinfo {author} {\bibfnamefont {B.}~\bibnamefont {Scheiner}},
  \bibinfo {author} {\bibfnamefont {B.~T.}\ \bibnamefont {Yee}}, \bibinfo
  {author} {\bibfnamefont {M.~M.}\ \bibnamefont {Hopkins}},\ and\ \bibinfo
  {author} {\bibfnamefont {E.}~\bibnamefont {Barnat}},\ }\bibfield  {title}
  {\enquote {\bibinfo {title} {Interaction of biased electrodes and plasmas:
  sheaths, double layers, and fireballs},}\ }\href@noop {} {\bibfield
  {journal} {\bibinfo  {journal} {Plasma Sources Science and Technology}\
  }\textbf {\bibinfo {volume} {29}},\ \bibinfo {pages} {053001} (\bibinfo
  {year} {2020})}\BibitemShut {NoStop}%
\bibitem [{\citenamefont {Robertson}(2013)}]{robertson2013sheaths}%
  \BibitemOpen
  \bibfield  {author} {\bibinfo {author} {\bibfnamefont {S.}~\bibnamefont
  {Robertson}},\ }\bibfield  {title} {\enquote {\bibinfo {title} {Sheaths in
  laboratory and space plasmas},}\ }\href@noop {} {\bibfield  {journal}
  {\bibinfo  {journal} {Plasma Physics and Controlled Fusion}\ }\textbf
  {\bibinfo {volume} {55}},\ \bibinfo {pages} {093001} (\bibinfo {year}
  {2013})}\BibitemShut {NoStop}%
\bibitem [{\citenamefont {Wollenhaupt}, \citenamefont {Le},\ and\ \citenamefont
  {Herdrich}(2018)}]{wollenhaupt2018overview}%
  \BibitemOpen
  \bibfield  {author} {\bibinfo {author} {\bibfnamefont {B.}~\bibnamefont
  {Wollenhaupt}}, \bibinfo {author} {\bibfnamefont {Q.~H.}\ \bibnamefont
  {Le}},\ and\ \bibinfo {author} {\bibfnamefont {G.}~\bibnamefont {Herdrich}},\
  }\bibfield  {title} {\enquote {\bibinfo {title} {Overview of thermal arcjet
  thruster development},}\ }\href@noop {} {\bibfield  {journal} {\bibinfo
  {journal} {Aircraft Engineering and Aerospace Technology}\ } (\bibinfo {year}
  {2018})}\BibitemShut {NoStop}%
\bibitem [{\citenamefont {Giuliani}\ and\ \citenamefont
  {Commisso}(2015)}]{giuliani2015review}%
  \BibitemOpen
  \bibfield  {author} {\bibinfo {author} {\bibfnamefont {J.~L.}\ \bibnamefont
  {Giuliani}}\ and\ \bibinfo {author} {\bibfnamefont {R.~J.}\ \bibnamefont
  {Commisso}},\ }\bibfield  {title} {\enquote {\bibinfo {title} {A review of
  the gas-puff $ z $-pinch as an x-ray and neutron source},}\ }\href@noop {}
  {\bibfield  {journal} {\bibinfo  {journal} {IEEE Transactions on Plasma
  Science}\ }\textbf {\bibinfo {volume} {43}},\ \bibinfo {pages} {2385--2453}
  (\bibinfo {year} {2015})}\BibitemShut {NoStop}%
\bibitem [{\citenamefont {Weynants}\ and\ \citenamefont
  {Van~Oost}(1993)}]{weynants1993edge}%
  \BibitemOpen
  \bibfield  {author} {\bibinfo {author} {\bibfnamefont {R.}~\bibnamefont
  {Weynants}}\ and\ \bibinfo {author} {\bibfnamefont {G.}~\bibnamefont
  {Van~Oost}},\ }\bibfield  {title} {\enquote {\bibinfo {title} {Edge biasing
  in tokamaks},}\ }\href@noop {} {\bibfield  {journal} {\bibinfo  {journal}
  {Plasma physics and controlled fusion}\ }\textbf {\bibinfo {volume} {35}},\
  \bibinfo {pages} {B177} (\bibinfo {year} {1993})}\BibitemShut {NoStop}%
\bibitem [{\citenamefont {Shumlak}(2020)}]{shumlak2020z}%
  \BibitemOpen
  \bibfield  {author} {\bibinfo {author} {\bibfnamefont {U.}~\bibnamefont
  {Shumlak}},\ }\bibfield  {title} {\enquote {\bibinfo {title} {Z-pinch
  fusion},}\ }\href@noop {} {\bibfield  {journal} {\bibinfo  {journal} {Journal
  of Applied Physics}\ }\textbf {\bibinfo {volume} {127}},\ \bibinfo {pages}
  {200901} (\bibinfo {year} {2020})}\BibitemShut {NoStop}%
\bibitem [{\citenamefont {Hartman}\ \emph {et~al.}(1977)\citenamefont
  {Hartman}, \citenamefont {Carlson}, \citenamefont {Hoffman}, \citenamefont
  {Werner},\ and\ \citenamefont {Cheng}}]{hartman1977conceptual}%
  \BibitemOpen
  \bibfield  {author} {\bibinfo {author} {\bibfnamefont {C.}~\bibnamefont
  {Hartman}}, \bibinfo {author} {\bibfnamefont {G.}~\bibnamefont {Carlson}},
  \bibinfo {author} {\bibfnamefont {M.}~\bibnamefont {Hoffman}}, \bibinfo
  {author} {\bibfnamefont {R.}~\bibnamefont {Werner}},\ and\ \bibinfo {author}
  {\bibfnamefont {D.}~\bibnamefont {Cheng}},\ }\bibfield  {title} {\enquote
  {\bibinfo {title} {A conceptual fusion reactor based on the
  high-plasma-density z-pinch},}\ }\href@noop {} {\bibfield  {journal}
  {\bibinfo  {journal} {Nuclear Fusion}\ }\textbf {\bibinfo {volume} {17}},\
  \bibinfo {pages} {909} (\bibinfo {year} {1977})}\BibitemShut {NoStop}%
\bibitem [{\citenamefont {Bennett}(1934)}]{bennett1934magnetically}%
  \BibitemOpen
  \bibfield  {author} {\bibinfo {author} {\bibfnamefont {W.~H.}\ \bibnamefont
  {Bennett}},\ }\bibfield  {title} {\enquote {\bibinfo {title} {Magnetically
  self-focussing streams},}\ }\href@noop {} {\bibfield  {journal} {\bibinfo
  {journal} {Physical Review}\ }\textbf {\bibinfo {volume} {45}},\ \bibinfo
  {pages} {890} (\bibinfo {year} {1934})}\BibitemShut {NoStop}%
\bibitem [{\citenamefont {Haines}\ \emph {et~al.}(2000)\citenamefont {Haines},
  \citenamefont {Lebedev}, \citenamefont {Chittenden}, \citenamefont {Beg},
  \citenamefont {Bland},\ and\ \citenamefont {Dangor}}]{haines2000past}%
  \BibitemOpen
  \bibfield  {author} {\bibinfo {author} {\bibfnamefont {M.}~\bibnamefont
  {Haines}}, \bibinfo {author} {\bibfnamefont {S.}~\bibnamefont {Lebedev}},
  \bibinfo {author} {\bibfnamefont {J.}~\bibnamefont {Chittenden}}, \bibinfo
  {author} {\bibfnamefont {F.}~\bibnamefont {Beg}}, \bibinfo {author}
  {\bibfnamefont {S.}~\bibnamefont {Bland}},\ and\ \bibinfo {author}
  {\bibfnamefont {A.}~\bibnamefont {Dangor}},\ }\bibfield  {title} {\enquote
  {\bibinfo {title} {The past, present, and future of z pinches},}\ }\href@noop
  {} {\bibfield  {journal} {\bibinfo  {journal} {Physics of Plasmas}\ }\textbf
  {\bibinfo {volume} {7}},\ \bibinfo {pages} {1672--1680} (\bibinfo {year}
  {2000})}\BibitemShut {NoStop}%
\bibitem [{\citenamefont {Shumlak}\ \emph {et~al.}(2001)\citenamefont
  {Shumlak}, \citenamefont {Golingo}, \citenamefont {Nelson},\ and\
  \citenamefont {Den~Hartog}}]{shumlak2001evidence}%
  \BibitemOpen
  \bibfield  {author} {\bibinfo {author} {\bibfnamefont {U.}~\bibnamefont
  {Shumlak}}, \bibinfo {author} {\bibfnamefont {R.}~\bibnamefont {Golingo}},
  \bibinfo {author} {\bibfnamefont {B.}~\bibnamefont {Nelson}},\ and\ \bibinfo
  {author} {\bibfnamefont {D.}~\bibnamefont {Den~Hartog}},\ }\bibfield  {title}
  {\enquote {\bibinfo {title} {Evidence of stabilization in the z-pinch},}\
  }\href@noop {} {\bibfield  {journal} {\bibinfo  {journal} {Physical review
  letters}\ }\textbf {\bibinfo {volume} {87}},\ \bibinfo {pages} {205005}
  (\bibinfo {year} {2001})}\BibitemShut {NoStop}%
\bibitem [{\citenamefont {Shumlak}\ \emph {et~al.}(2003)\citenamefont
  {Shumlak}, \citenamefont {Nelson}, \citenamefont {Golingo}, \citenamefont
  {Jackson}, \citenamefont {Crawford},\ and\ \citenamefont
  {Den~Hartog}}]{shumlak2003sheared}%
  \BibitemOpen
  \bibfield  {author} {\bibinfo {author} {\bibfnamefont {U.}~\bibnamefont
  {Shumlak}}, \bibinfo {author} {\bibfnamefont {B.}~\bibnamefont {Nelson}},
  \bibinfo {author} {\bibfnamefont {R.}~\bibnamefont {Golingo}}, \bibinfo
  {author} {\bibfnamefont {S.}~\bibnamefont {Jackson}}, \bibinfo {author}
  {\bibfnamefont {E.}~\bibnamefont {Crawford}},\ and\ \bibinfo {author}
  {\bibfnamefont {D.}~\bibnamefont {Den~Hartog}},\ }\bibfield  {title}
  {\enquote {\bibinfo {title} {Sheared flow stabilization experiments in the
  zap flow z pinch},}\ }\href@noop {} {\bibfield  {journal} {\bibinfo
  {journal} {Physics of Plasmas}\ }\textbf {\bibinfo {volume} {10}},\ \bibinfo
  {pages} {1683--1690} (\bibinfo {year} {2003})}\BibitemShut {NoStop}%
\bibitem [{\citenamefont {Shumlak}\ \emph {et~al.}(2009)\citenamefont
  {Shumlak}, \citenamefont {Adams}, \citenamefont {Blakely}, \citenamefont
  {Chan}, \citenamefont {Golingo}, \citenamefont {Knecht}, \citenamefont
  {Nelson}, \citenamefont {Oberto}, \citenamefont {Sybouts},\ and\
  \citenamefont {Vogman}}]{shumlak2009equilibrium}%
  \BibitemOpen
  \bibfield  {author} {\bibinfo {author} {\bibfnamefont {U.}~\bibnamefont
  {Shumlak}}, \bibinfo {author} {\bibfnamefont {C.}~\bibnamefont {Adams}},
  \bibinfo {author} {\bibfnamefont {J.}~\bibnamefont {Blakely}}, \bibinfo
  {author} {\bibfnamefont {B.-J.}\ \bibnamefont {Chan}}, \bibinfo {author}
  {\bibfnamefont {R.}~\bibnamefont {Golingo}}, \bibinfo {author} {\bibfnamefont
  {S.}~\bibnamefont {Knecht}}, \bibinfo {author} {\bibfnamefont
  {B.}~\bibnamefont {Nelson}}, \bibinfo {author} {\bibfnamefont
  {R.}~\bibnamefont {Oberto}}, \bibinfo {author} {\bibfnamefont
  {M.}~\bibnamefont {Sybouts}},\ and\ \bibinfo {author} {\bibfnamefont
  {G.}~\bibnamefont {Vogman}},\ }\bibfield  {title} {\enquote {\bibinfo {title}
  {Equilibrium, flow shear and stability measurements in the z-pinch},}\
  }\href@noop {} {\bibfield  {journal} {\bibinfo  {journal} {Nuclear Fusion}\
  }\textbf {\bibinfo {volume} {49}},\ \bibinfo {pages} {075039} (\bibinfo
  {year} {2009})}\BibitemShut {NoStop}%
\bibitem [{\citenamefont {Shumlak}\ \emph {et~al.}(2017)\citenamefont
  {Shumlak}, \citenamefont {Nelson}, \citenamefont {Claveau}, \citenamefont
  {Forbes}, \citenamefont {Golingo}, \citenamefont {Hughes}, \citenamefont
  {Oberto}, \citenamefont {Ross},\ and\ \citenamefont
  {Weber}}]{shumlak2017increasing}%
  \BibitemOpen
  \bibfield  {author} {\bibinfo {author} {\bibfnamefont {U.}~\bibnamefont
  {Shumlak}}, \bibinfo {author} {\bibfnamefont {B.}~\bibnamefont {Nelson}},
  \bibinfo {author} {\bibfnamefont {E.}~\bibnamefont {Claveau}}, \bibinfo
  {author} {\bibfnamefont {E.}~\bibnamefont {Forbes}}, \bibinfo {author}
  {\bibfnamefont {R.}~\bibnamefont {Golingo}}, \bibinfo {author} {\bibfnamefont
  {M.}~\bibnamefont {Hughes}}, \bibinfo {author} {\bibfnamefont
  {R.}~\bibnamefont {Oberto}}, \bibinfo {author} {\bibfnamefont
  {M.}~\bibnamefont {Ross}},\ and\ \bibinfo {author} {\bibfnamefont
  {T.}~\bibnamefont {Weber}},\ }\bibfield  {title} {\enquote {\bibinfo {title}
  {Increasing plasma parameters using sheared flow stabilization of a
  z-pinch},}\ }\href@noop {} {\bibfield  {journal} {\bibinfo  {journal}
  {Physics of Plasmas}\ }\textbf {\bibinfo {volume} {24}},\ \bibinfo {pages}
  {055702} (\bibinfo {year} {2017})}\BibitemShut {NoStop}%
\bibitem [{\citenamefont {Zhang}\ \emph {et~al.}(2019)\citenamefont {Zhang},
  \citenamefont {Shumlak}, \citenamefont {Nelson}, \citenamefont {Golingo},
  \citenamefont {Weber}, \citenamefont {Stepanov}, \citenamefont {Claveau},
  \citenamefont {Forbes}, \citenamefont {Draper}, \citenamefont {Mitrani} \emph
  {et~al.}}]{zhang2019sustained}%
  \BibitemOpen
  \bibfield  {author} {\bibinfo {author} {\bibfnamefont {Y.}~\bibnamefont
  {Zhang}}, \bibinfo {author} {\bibfnamefont {U.}~\bibnamefont {Shumlak}},
  \bibinfo {author} {\bibfnamefont {B.}~\bibnamefont {Nelson}}, \bibinfo
  {author} {\bibfnamefont {R.}~\bibnamefont {Golingo}}, \bibinfo {author}
  {\bibfnamefont {T.}~\bibnamefont {Weber}}, \bibinfo {author} {\bibfnamefont
  {A.}~\bibnamefont {Stepanov}}, \bibinfo {author} {\bibfnamefont
  {E.}~\bibnamefont {Claveau}}, \bibinfo {author} {\bibfnamefont
  {E.}~\bibnamefont {Forbes}}, \bibinfo {author} {\bibfnamefont
  {Z.}~\bibnamefont {Draper}}, \bibinfo {author} {\bibfnamefont
  {J.}~\bibnamefont {Mitrani}}, \emph {et~al.},\ }\bibfield  {title} {\enquote
  {\bibinfo {title} {Sustained neutron production from a sheared-flow
  stabilized z pinch},}\ }\href@noop {} {\bibfield  {journal} {\bibinfo
  {journal} {Physical review letters}\ }\textbf {\bibinfo {volume} {122}},\
  \bibinfo {pages} {135001} (\bibinfo {year} {2019})}\BibitemShut {NoStop}%
\bibitem [{\citenamefont {Mitrani}\ \emph {et~al.}(2019)\citenamefont
  {Mitrani}, \citenamefont {Higginson}, \citenamefont {Draper}, \citenamefont
  {Morrell}, \citenamefont {Bernstein}, \citenamefont {Claveau}, \citenamefont
  {Cooper}, \citenamefont {Forbes}, \citenamefont {Golingo}, \citenamefont
  {Nelson} \emph {et~al.}}]{mitrani2019measurements}%
  \BibitemOpen
  \bibfield  {author} {\bibinfo {author} {\bibfnamefont {J.~M.}\ \bibnamefont
  {Mitrani}}, \bibinfo {author} {\bibfnamefont {D.~P.}\ \bibnamefont
  {Higginson}}, \bibinfo {author} {\bibfnamefont {Z.~T.}\ \bibnamefont
  {Draper}}, \bibinfo {author} {\bibfnamefont {J.}~\bibnamefont {Morrell}},
  \bibinfo {author} {\bibfnamefont {L.~A.}\ \bibnamefont {Bernstein}}, \bibinfo
  {author} {\bibfnamefont {E.~L.}\ \bibnamefont {Claveau}}, \bibinfo {author}
  {\bibfnamefont {C.~M.}\ \bibnamefont {Cooper}}, \bibinfo {author}
  {\bibfnamefont {E.~G.}\ \bibnamefont {Forbes}}, \bibinfo {author}
  {\bibfnamefont {R.~P.}\ \bibnamefont {Golingo}}, \bibinfo {author}
  {\bibfnamefont {B.~A.}\ \bibnamefont {Nelson}}, \emph {et~al.},\ }\bibfield
  {title} {\enquote {\bibinfo {title} {Measurements of temporally-and
  spatially-resolved neutron production in a sheared-flow stabilized
  z-pinch},}\ }\href@noop {} {\bibfield  {journal} {\bibinfo  {journal}
  {Nuclear Instruments and Methods in Physics Research Section A: Accelerators,
  Spectrometers, Detectors and Associated Equipment}\ }\textbf {\bibinfo
  {volume} {947}},\ \bibinfo {pages} {162764} (\bibinfo {year}
  {2019})}\BibitemShut {NoStop}%
\bibitem [{\citenamefont {Claveau}\ \emph {et~al.}(2020)\citenamefont
  {Claveau}, \citenamefont {Shumlak}, \citenamefont {Nelson}, \citenamefont
  {Forbes}, \citenamefont {Stepanov}, \citenamefont {Weber}, \citenamefont
  {Zhang},\ and\ \citenamefont {McLean}}]{claveau2020plasma}%
  \BibitemOpen
  \bibfield  {author} {\bibinfo {author} {\bibfnamefont {E.}~\bibnamefont
  {Claveau}}, \bibinfo {author} {\bibfnamefont {U.}~\bibnamefont {Shumlak}},
  \bibinfo {author} {\bibfnamefont {B.}~\bibnamefont {Nelson}}, \bibinfo
  {author} {\bibfnamefont {E.}~\bibnamefont {Forbes}}, \bibinfo {author}
  {\bibfnamefont {A.}~\bibnamefont {Stepanov}}, \bibinfo {author}
  {\bibfnamefont {T.}~\bibnamefont {Weber}}, \bibinfo {author} {\bibfnamefont
  {Y.}~\bibnamefont {Zhang}},\ and\ \bibinfo {author} {\bibfnamefont
  {H.}~\bibnamefont {McLean}},\ }\bibfield  {title} {\enquote {\bibinfo {title}
  {Plasma exhaust in a sheared-flow-stabilized z pinch},}\ }\href@noop {}
  {\bibfield  {journal} {\bibinfo  {journal} {Physics of Plasmas}\ }\textbf
  {\bibinfo {volume} {27}},\ \bibinfo {pages} {092510} (\bibinfo {year}
  {2020})}\BibitemShut {NoStop}%
\bibitem [{\citenamefont {Stepanov}\ \emph {et~al.}(2020)\citenamefont
  {Stepanov}, \citenamefont {Shumlak}, \citenamefont {McLean}, \citenamefont
  {Nelson}, \citenamefont {Claveau}, \citenamefont {Forbes}, \citenamefont
  {Weber},\ and\ \citenamefont {Zhang}}]{stepanov2020flow}%
  \BibitemOpen
  \bibfield  {author} {\bibinfo {author} {\bibfnamefont {A.}~\bibnamefont
  {Stepanov}}, \bibinfo {author} {\bibfnamefont {U.}~\bibnamefont {Shumlak}},
  \bibinfo {author} {\bibfnamefont {H.}~\bibnamefont {McLean}}, \bibinfo
  {author} {\bibfnamefont {B.}~\bibnamefont {Nelson}}, \bibinfo {author}
  {\bibfnamefont {E.}~\bibnamefont {Claveau}}, \bibinfo {author} {\bibfnamefont
  {E.}~\bibnamefont {Forbes}}, \bibinfo {author} {\bibfnamefont
  {T.}~\bibnamefont {Weber}},\ and\ \bibinfo {author} {\bibfnamefont
  {Y.}~\bibnamefont {Zhang}},\ }\bibfield  {title} {\enquote {\bibinfo {title}
  {Flow z-pinch plasma production on the fuze experiment},}\ }\href@noop {}
  {\bibfield  {journal} {\bibinfo  {journal} {Physics of Plasmas}\ }\textbf
  {\bibinfo {volume} {27}},\ \bibinfo {pages} {112503} (\bibinfo {year}
  {2020})}\BibitemShut {NoStop}%
\bibitem [{\citenamefont {Shumlak}\ \emph {et~al.}(2012)\citenamefont
  {Shumlak}, \citenamefont {Chadney}, \citenamefont {Golingo}, \citenamefont
  {Den~Hartog}, \citenamefont {Hughes}, \citenamefont {Knecht}, \citenamefont
  {Lowrie}, \citenamefont {Lukin}, \citenamefont {Nelson}, \citenamefont
  {Oberto} \emph {et~al.}}]{shumlak2012sheared}%
  \BibitemOpen
  \bibfield  {author} {\bibinfo {author} {\bibfnamefont {U.}~\bibnamefont
  {Shumlak}}, \bibinfo {author} {\bibfnamefont {J.}~\bibnamefont {Chadney}},
  \bibinfo {author} {\bibfnamefont {R.}~\bibnamefont {Golingo}}, \bibinfo
  {author} {\bibfnamefont {D.}~\bibnamefont {Den~Hartog}}, \bibinfo {author}
  {\bibfnamefont {M.}~\bibnamefont {Hughes}}, \bibinfo {author} {\bibfnamefont
  {S.}~\bibnamefont {Knecht}}, \bibinfo {author} {\bibfnamefont
  {W.}~\bibnamefont {Lowrie}}, \bibinfo {author} {\bibfnamefont
  {V.}~\bibnamefont {Lukin}}, \bibinfo {author} {\bibfnamefont
  {B.}~\bibnamefont {Nelson}}, \bibinfo {author} {\bibfnamefont
  {R.}~\bibnamefont {Oberto}}, \emph {et~al.},\ }\bibfield  {title} {\enquote
  {\bibinfo {title} {The sheared-flow stabilized z-pinch},}\ }\href@noop {}
  {\bibfield  {journal} {\bibinfo  {journal} {Fusion Science and Technology}\
  }\textbf {\bibinfo {volume} {61}},\ \bibinfo {pages} {119--124} (\bibinfo
  {year} {2012})}\BibitemShut {NoStop}%
\bibitem [{\citenamefont {Scheiner}\ \emph {et~al.}(2016)\citenamefont
  {Scheiner}, \citenamefont {Baalrud}, \citenamefont {Hopkins}, \citenamefont
  {Yee},\ and\ \citenamefont {Barnat}}]{scheiner2016particle}%
  \BibitemOpen
  \bibfield  {author} {\bibinfo {author} {\bibfnamefont {B.}~\bibnamefont
  {Scheiner}}, \bibinfo {author} {\bibfnamefont {S.~D.}\ \bibnamefont
  {Baalrud}}, \bibinfo {author} {\bibfnamefont {M.~M.}\ \bibnamefont
  {Hopkins}}, \bibinfo {author} {\bibfnamefont {B.~T.}\ \bibnamefont {Yee}},\
  and\ \bibinfo {author} {\bibfnamefont {E.~V.}\ \bibnamefont {Barnat}},\
  }\bibfield  {title} {\enquote {\bibinfo {title} {Particle-in-cell study of
  the ion-to-electron sheath transition},}\ }\href@noop {} {\bibfield
  {journal} {\bibinfo  {journal} {Physics of Plasmas}\ }\textbf {\bibinfo
  {volume} {23}},\ \bibinfo {pages} {083510} (\bibinfo {year}
  {2016})}\BibitemShut {NoStop}%
\bibitem [{\citenamefont {Li}\ \emph {et~al.}(2022{\natexlab{a}})\citenamefont
  {Li}, \citenamefont {Srinivasan}, \citenamefont {Zhang},\ and\ \citenamefont
  {Tang}}]{li2022bohm}%
  \BibitemOpen
  \bibfield  {author} {\bibinfo {author} {\bibfnamefont {Y.}~\bibnamefont
  {Li}}, \bibinfo {author} {\bibfnamefont {B.}~\bibnamefont {Srinivasan}},
  \bibinfo {author} {\bibfnamefont {Y.}~\bibnamefont {Zhang}},\ and\ \bibinfo
  {author} {\bibfnamefont {X.-Z.}\ \bibnamefont {Tang}},\ }\bibfield  {title}
  {\enquote {\bibinfo {title} {Bohm criterion of plasma sheaths away from
  asymptotic limits},}\ }\href@noop {} {\bibfield  {journal} {\bibinfo
  {journal} {Physical Review Letters}\ }\textbf {\bibinfo {volume} {128}},\
  \bibinfo {pages} {085002} (\bibinfo {year} {2022}{\natexlab{a}})}\BibitemShut
  {NoStop}%
\bibitem [{\citenamefont {Li}\ \emph {et~al.}(2022{\natexlab{b}})\citenamefont
  {Li}, \citenamefont {Srinivasan}, \citenamefont {Zhang},\ and\ \citenamefont
  {Tang}}]{li2022transport}%
  \BibitemOpen
  \bibfield  {author} {\bibinfo {author} {\bibfnamefont {Y.}~\bibnamefont
  {Li}}, \bibinfo {author} {\bibfnamefont {B.}~\bibnamefont {Srinivasan}},
  \bibinfo {author} {\bibfnamefont {Y.}~\bibnamefont {Zhang}},\ and\ \bibinfo
  {author} {\bibfnamefont {X.-Z.}\ \bibnamefont {Tang}},\ }\bibfield  {title}
  {\enquote {\bibinfo {title} {Transport physics dependence of bohm speed in
  presheath--sheath transition},}\ }\href@noop {} {\bibfield  {journal}
  {\bibinfo  {journal} {Physics of Plasmas}\ }\textbf {\bibinfo {volume}
  {29}},\ \bibinfo {pages} {113509} (\bibinfo {year}
  {2022}{\natexlab{b}})}\BibitemShut {NoStop}%
\bibitem [{\citenamefont {Shumlak}\ \emph {et~al.}(2011)\citenamefont
  {Shumlak}, \citenamefont {Lilly}, \citenamefont {Reddell}, \citenamefont
  {Sousa},\ and\ \citenamefont {Srinivasan}}]{shumlak2011advanced}%
  \BibitemOpen
  \bibfield  {author} {\bibinfo {author} {\bibfnamefont {U.}~\bibnamefont
  {Shumlak}}, \bibinfo {author} {\bibfnamefont {R.}~\bibnamefont {Lilly}},
  \bibinfo {author} {\bibfnamefont {N.}~\bibnamefont {Reddell}}, \bibinfo
  {author} {\bibfnamefont {E.}~\bibnamefont {Sousa}},\ and\ \bibinfo {author}
  {\bibfnamefont {B.}~\bibnamefont {Srinivasan}},\ }\bibfield  {title}
  {\enquote {\bibinfo {title} {Advanced physics calculations using a
  multi-fluid plasma model},}\ }\href@noop {} {\bibfield  {journal} {\bibinfo
  {journal} {Computer Physics Communications}\ }\textbf {\bibinfo {volume}
  {182}},\ \bibinfo {pages} {1767--1770} (\bibinfo {year} {2011})}\BibitemShut
  {NoStop}%
\bibitem [{\citenamefont {Cagas}\ \emph {et~al.}(2017)\citenamefont {Cagas},
  \citenamefont {Hakim}, \citenamefont {Juno},\ and\ \citenamefont
  {Srinivasan}}]{cagas2017continuum}%
  \BibitemOpen
  \bibfield  {author} {\bibinfo {author} {\bibfnamefont {P.}~\bibnamefont
  {Cagas}}, \bibinfo {author} {\bibfnamefont {A.}~\bibnamefont {Hakim}},
  \bibinfo {author} {\bibfnamefont {J.}~\bibnamefont {Juno}},\ and\ \bibinfo
  {author} {\bibfnamefont {B.}~\bibnamefont {Srinivasan}},\ }\bibfield  {title}
  {\enquote {\bibinfo {title} {Continuum kinetic and multi-fluid simulations of
  classical sheaths},}\ }\href@noop {} {\bibfield  {journal} {\bibinfo
  {journal} {Physics of Plasmas}\ }\textbf {\bibinfo {volume} {24}},\ \bibinfo
  {pages} {022118} (\bibinfo {year} {2017})}\BibitemShut {NoStop}%
\bibitem [{\citenamefont {Cagas}, \citenamefont {Hakim},\ and\ \citenamefont
  {Srinivasan}(2020)}]{cagas2020plasma}%
  \BibitemOpen
  \bibfield  {author} {\bibinfo {author} {\bibfnamefont {P.}~\bibnamefont
  {Cagas}}, \bibinfo {author} {\bibfnamefont {A.}~\bibnamefont {Hakim}},\ and\
  \bibinfo {author} {\bibfnamefont {B.}~\bibnamefont {Srinivasan}},\ }\bibfield
   {title} {\enquote {\bibinfo {title} {Plasma-material boundary conditions for
  discontinuous galerkin continuum-kinetic simulations, with a focus on
  secondary electron emission},}\ }\href@noop {} {\bibfield  {journal}
  {\bibinfo  {journal} {Journal of Computational Physics}\ }\textbf {\bibinfo
  {volume} {406}},\ \bibinfo {pages} {109215} (\bibinfo {year}
  {2020})}\BibitemShut {NoStop}%
\bibitem [{\citenamefont {Bradshaw}\ \emph {et~al.}(2022)\citenamefont
  {Bradshaw}, \citenamefont {Cagas}, \citenamefont {Hakim},\ and\ \citenamefont
  {Srinivasan}}]{bradshaw2022}%
  \BibitemOpen
  \bibfield  {author} {\bibinfo {author} {\bibfnamefont {K.}~\bibnamefont
  {Bradshaw}}, \bibinfo {author} {\bibfnamefont {P.}~\bibnamefont {Cagas}},
  \bibinfo {author} {\bibfnamefont {A.}~\bibnamefont {Hakim}},\ and\ \bibinfo
  {author} {\bibfnamefont {B.}~\bibnamefont {Srinivasan}},\ }\href
  {https://doi.org/10.48550/ARXIV.2210.14117} {\enquote {\bibinfo {title}
  {Plasma sheath studies using a physical treatment of electron emission from a
  dielectric wall},}\ } (\bibinfo {year} {2022})\BibitemShut {NoStop}%
\bibitem [{\citenamefont {Dougherty}(1964)}]{dougherty1964model}%
  \BibitemOpen
  \bibfield  {author} {\bibinfo {author} {\bibfnamefont {J.}~\bibnamefont
  {Dougherty}},\ }\bibfield  {title} {\enquote {\bibinfo {title} {Model
  fokker-planck equation for a plasma and its solution},}\ }\href@noop {}
  {\bibfield  {journal} {\bibinfo  {journal} {The Physics of Fluids}\ }\textbf
  {\bibinfo {volume} {7}},\ \bibinfo {pages} {1788--1799} (\bibinfo {year}
  {1964})}\BibitemShut {NoStop}%
\bibitem [{\citenamefont {Francisquez}\ \emph {et~al.}(2020)\citenamefont
  {Francisquez}, \citenamefont {Bernard}, \citenamefont {Mandell},
  \citenamefont {Hammett},\ and\ \citenamefont
  {Hakim}}]{francisquez2020conservative}%
  \BibitemOpen
  \bibfield  {author} {\bibinfo {author} {\bibfnamefont {M.}~\bibnamefont
  {Francisquez}}, \bibinfo {author} {\bibfnamefont {T.~N.}\ \bibnamefont
  {Bernard}}, \bibinfo {author} {\bibfnamefont {N.~R.}\ \bibnamefont
  {Mandell}}, \bibinfo {author} {\bibfnamefont {G.~W.}\ \bibnamefont
  {Hammett}},\ and\ \bibinfo {author} {\bibfnamefont {A.}~\bibnamefont
  {Hakim}},\ }\bibfield  {title} {\enquote {\bibinfo {title} {Conservative
  discontinuous galerkin scheme of a gyro-averaged dougherty collision
  operator},}\ }\href@noop {} {\bibfield  {journal} {\bibinfo  {journal}
  {Nuclear Fusion}\ }\textbf {\bibinfo {volume} {60}},\ \bibinfo {pages}
  {096021} (\bibinfo {year} {2020})}\BibitemShut {NoStop}%
\bibitem [{\citenamefont {Hakim}\ \emph {et~al.}(2020)\citenamefont {Hakim},
  \citenamefont {Francisquez}, \citenamefont {Juno},\ and\ \citenamefont
  {Hammett}}]{hakim2020conservative}%
  \BibitemOpen
  \bibfield  {author} {\bibinfo {author} {\bibfnamefont {A.}~\bibnamefont
  {Hakim}}, \bibinfo {author} {\bibfnamefont {M.}~\bibnamefont {Francisquez}},
  \bibinfo {author} {\bibfnamefont {J.}~\bibnamefont {Juno}},\ and\ \bibinfo
  {author} {\bibfnamefont {G.~W.}\ \bibnamefont {Hammett}},\ }\bibfield
  {title} {\enquote {\bibinfo {title} {Conservative discontinuous galerkin
  schemes for nonlinear dougherty--fokker--planck collision operators},}\
  }\href@noop {} {\bibfield  {journal} {\bibinfo  {journal} {Journal of Plasma
  Physics}\ }\textbf {\bibinfo {volume} {86}} (\bibinfo {year}
  {2020})}\BibitemShut {NoStop}%
\bibitem [{\citenamefont {Wang}\ \emph {et~al.}(2015)\citenamefont {Wang},
  \citenamefont {Hakim}, \citenamefont {Bhattacharjee},\ and\ \citenamefont
  {Germaschewski}}]{wangL2015}%
  \BibitemOpen
  \bibfield  {author} {\bibinfo {author} {\bibfnamefont {L.}~\bibnamefont
  {Wang}}, \bibinfo {author} {\bibfnamefont {A.~H.}\ \bibnamefont {Hakim}},
  \bibinfo {author} {\bibfnamefont {A.}~\bibnamefont {Bhattacharjee}},\ and\
  \bibinfo {author} {\bibfnamefont {K.}~\bibnamefont {Germaschewski}},\
  }\bibfield  {title} {\enquote {\bibinfo {title} {Comparison of multi-fluid
  moment models with particle-in-cell simulations of collisionless magnetic
  reconnection},}\ }\href@noop {} {\bibfield  {journal} {\bibinfo  {journal}
  {Physics of Plasmas}\ }\textbf {\bibinfo {volume} {22}},\ \bibinfo {pages}
  {012108} (\bibinfo {year} {2015})}\BibitemShut {NoStop}%
\bibitem [{\citenamefont {Cagas}(2018)}]{cagas2018continuum}%
  \BibitemOpen
  \bibfield  {author} {\bibinfo {author} {\bibfnamefont {P.}~\bibnamefont
  {Cagas}},\ }\emph {\bibinfo {title} {Continuum Kinetic Simulations of Plasma
  Sheaths and Instabilities}},\ \href@noop {} {Ph.D. thesis},\ \bibinfo
  {school} {Virginia Polytechnic Institute and State University} (\bibinfo
  {year} {2018})\BibitemShut {NoStop}%
\bibitem [{\citenamefont {Gkeyll}(2022)}]{gkylDocs}%
  \BibitemOpen
  \bibfield  {author} {\bibinfo {author} {\bibnamefont {Gkeyll}},\ }\href@noop
  {} {} (\bibinfo {year} {2022}),\ \bibinfo {note}
  {\url{https://gkeyll.readthedocs.io}}\BibitemShut {NoStop}%
\bibitem [{\citenamefont {Juno}\ \emph {et~al.}(2018)\citenamefont {Juno},
  \citenamefont {Hakim}, \citenamefont {TenBarge}, \citenamefont {Shi},\ and\
  \citenamefont {Dorland}}]{juno2018discontinuous}%
  \BibitemOpen
  \bibfield  {author} {\bibinfo {author} {\bibfnamefont {J.}~\bibnamefont
  {Juno}}, \bibinfo {author} {\bibfnamefont {A.}~\bibnamefont {Hakim}},
  \bibinfo {author} {\bibfnamefont {J.}~\bibnamefont {TenBarge}}, \bibinfo
  {author} {\bibfnamefont {E.}~\bibnamefont {Shi}},\ and\ \bibinfo {author}
  {\bibfnamefont {W.}~\bibnamefont {Dorland}},\ }\bibfield  {title} {\enquote
  {\bibinfo {title} {Discontinuous galerkin algorithms for fully kinetic
  plasmas},}\ }\href@noop {} {\bibfield  {journal} {\bibinfo  {journal}
  {Journal of Computational Physics}\ }\textbf {\bibinfo {volume} {353}},\
  \bibinfo {pages} {110--147} (\bibinfo {year} {2018})}\BibitemShut {NoStop}%
\bibitem [{\citenamefont {Hakim}\ and\ \citenamefont
  {Juno}(2020)}]{hakim2020alias}%
  \BibitemOpen
  \bibfield  {author} {\bibinfo {author} {\bibfnamefont {A.}~\bibnamefont
  {Hakim}}\ and\ \bibinfo {author} {\bibfnamefont {J.}~\bibnamefont {Juno}},\
  }\bibfield  {title} {\enquote {\bibinfo {title} {Alias-free, matrix-free, and
  quadrature-free discontinuous galerkin algorithms for (plasma) kinetic
  equations},}\ }in\ \href@noop {} {\emph {\bibinfo {booktitle} {SC20:
  International Conference for High Performance Computing, Networking, Storage
  and Analysis}}}\ (\bibinfo {organization} {IEEE},\ \bibinfo {year} {2020})\
  pp.\ \bibinfo {pages} {1--15}\BibitemShut {NoStop}%
\bibitem [{\citenamefont {Gottlieb}(2005)}]{gottlieb2005high}%
  \BibitemOpen
  \bibfield  {author} {\bibinfo {author} {\bibfnamefont {S.}~\bibnamefont
  {Gottlieb}},\ }\bibfield  {title} {\enquote {\bibinfo {title} {On high order
  strong stability preserving runge-kutta and multi step time
  discretizations},}\ }\href@noop {} {\bibfield  {journal} {\bibinfo  {journal}
  {Journal of scientific computing}\ }\textbf {\bibinfo {volume} {25}},\
  \bibinfo {pages} {105--128} (\bibinfo {year} {2005})}\BibitemShut {NoStop}%
\bibitem [{\citenamefont {Zhang}\ \emph {et~al.}(2022)\citenamefont {Zhang},
  \citenamefont {Li}, \citenamefont {Srinivasan},\ and\ \citenamefont
  {Tang}}]{zhang2022resolving}%
  \BibitemOpen
  \bibfield  {author} {\bibinfo {author} {\bibfnamefont {Y.}~\bibnamefont
  {Zhang}}, \bibinfo {author} {\bibfnamefont {Y.}~\bibnamefont {Li}}, \bibinfo
  {author} {\bibfnamefont {B.}~\bibnamefont {Srinivasan}},\ and\ \bibinfo
  {author} {\bibfnamefont {X.-Z.}\ \bibnamefont {Tang}},\ }\bibfield  {title}
  {\enquote {\bibinfo {title} {Resolving the mystery of electron perpendicular
  temperature spike in the plasma sheath},}\ }\href@noop {} {\bibfield
  {journal} {\bibinfo  {journal} {arXiv preprint arXiv:2210.16711}\ } (\bibinfo
  {year} {2022})}\BibitemShut {NoStop}%
\bibitem [{Note1()}]{Note1}%
  \BibitemOpen
  \bibinfo {note}
  {Https://docs.scipy.org/doc/scipy/reference/generated/scipy.integrate.odeint.html}\BibitemShut
  {NoStop}%
\bibitem [{\citenamefont {Petzold}(1983)}]{petzold1983automatic}%
  \BibitemOpen
  \bibfield  {author} {\bibinfo {author} {\bibfnamefont {L.}~\bibnamefont
  {Petzold}},\ }\bibfield  {title} {\enquote {\bibinfo {title} {Automatic
  selection of methods for solving stiff and nonstiff systems of ordinary
  differential equations},}\ }\href@noop {} {\bibfield  {journal} {\bibinfo
  {journal} {SIAM journal on scientific and statistical computing}\ }\textbf
  {\bibinfo {volume} {4}},\ \bibinfo {pages} {136--148} (\bibinfo {year}
  {1983})}\BibitemShut {NoStop}%
\end{thebibliography}%

\end{document}